\documentclass[journal]{IEEEtran}
\usepackage{amsmath,amsfonts}
\usepackage{algorithmic}
\usepackage{algorithm}
\usepackage{array}
\usepackage{textcomp}
\usepackage{stfloats}
\usepackage{url}
\usepackage{verbatim}
\usepackage{graphicx}
\usepackage{setspace}%行距
\usepackage{multirow}%跨行
\usepackage{caption2}
\usepackage{psfrag}%编辑EPS图形中的字符包
\usepackage{epstopdf}
\usepackage{cite}{\tiny }
\usepackage{amsthm}
\usepackage{cuted}
\usepackage{subfigure}
\usepackage{color}
\usepackage{hyperref}
\hypersetup{
	hidelinks,
	colorlinks=true,
	linkcolor=blue,
	citecolor=blue
}
\theoremstyle{definition}
\newtheorem{thm}{Theorem}
\newtheorem{cor}{Corollary}
\newtheorem{lem}{Lemma}
\newtheorem{rem}{Remark}
\allowdisplaybreaks[4]
\begin{document}
	
	\title{Pinching Antenna-aided NOMA Systems with Internal Eavesdropping}
	
	\author{Haolian Chi,   
		Kunrui Cao, 
		Zhou Su, \IEEEmembership{Senior Member,~IEEE,}
		Lei Zhou, \\
		Panagiotis D. Diamantoulakis, \IEEEmembership{Senior Member,~IEEE,}
		Yuanwei Liu, \IEEEmembership{Fellow,~IEEE,}\\
		and George K. Karagiannidis, \IEEEmembership{Fellow,~IEEE}
		\thanks{Haolian Chi, Kunrui Cao,  and
			Lei Zhou are with Information Support Force Engineering University, Wuhan 430035, China, and also with the School of Information and Communications,  National University of Defense Technology, Wuhan 430035, China (e-mail: chihaolian20@nudt.edu.cn;  krcao@nudt.edu.cn;  cat{\textunderscore}radar@163.com).}
		\thanks{Zhou Su  is with the School of Cyber Science and Engineering, Xi'an Jiaotong University, Xi'an 710049, China (e-mail:
			zhousu@ieee.org).}
		\thanks{Panagiotis D. Diamantoulakis, and George K. Karagiannidis are with the Department of Electrical and Computer Engineering, Aristotle University of Thessaloniki, Thessaloniki 54124, Greece (e-mail: padiamant@gmail.com; geokarag@auth.gr).}
		\thanks{Yuanwei Liu is with the Department of Electrical and Electronic Engineering, the University of Hong Kong, Hong Kong, China (e-mail: yuan-wei@hku.hk).}
		% <-this % stops a space
	}
	
	%The paper headers
	%\markboth{Journal of \LaTeX\ Class Files,~Vol.~14, No.~8, August~2021}%
	%{Shell \MakeLowercase{\textit{et al.}}: A Sample Article Using IEEEtran.cls for IEEE Journals}
	%\IEEEpubid{0000--0000/00\$00.00~\copyright~2021 IEEE}
	% Remember, if you use this you must call \IEEEpubidadjcol in the second
	%column for its text to clear the IEEEpubid mark.
	
	\maketitle
	\vfill
	\begin{abstract}
		As a novel member of flexible antennas, the pinching antenna (PA) is realized by integrating small dielectric particles on a waveguide, offering unique regulatory capabilities on constructing line-of-sight (LoS) links and enhancing transceiver channels, reducing path loss and signal blockage. Meanwhile, non-orthogonal multiple access (NOMA) has become a potential technology of next-generation communications due to its remarkable advantages in spectrum efficiency and user access capability. The integration of PA and NOMA enables synergistic leveraging of PA’s channel regulation capability and NOMA’s multi-user multiplexing advantage, forming a complementary technical framework to deliver high-performance communication solutions.  
		However, the use of successive interference cancellation (SIC) introduces significant security risks to power-domain NOMA systems when internal eavesdropping is present. To this end,  this paper investigates the physical layer security of a PA-aided NOMA system where a nearby user is considered as an internal eavesdropper. We enhance the security of the NOMA system through optimizing the radiated power of PAs and analyze the secrecy performance by deriving the closed-form expressions for the secrecy outage probability (SOP). Furthermore, we extend the characterization of PA flexibility beyond deployment and scale adjustment to include flexible regulation of PA coupling length. Based on two conventional PA power models, i.e., the equal power model and the proportional power model, we propose a flexible power strategy to achieve secure transmission. The results highlight the potential of the PA-aided NOMA system in mitigating internal eavesdropping risks, and provide an effective strategy for optimizing power allocation and cell range of user activity.
	\end{abstract}
	\begin{IEEEkeywords}
		Pinching antenna, non-orthogonal multiple access, physical layer security,  internal eavesdropping.
	\end{IEEEkeywords}
	\section{Introduction}
	The sixth generation (6G) of wireless networks represents a revolutionary development forward in the evolution of wireless communication technology, transcending the limitations of the fifth generation (5G), to meet the  expanding and complex demands of emerging applications. 6G delivers ultra-high data rates, ultra-low latency, and ultra-high reliability, integrating multiple functions including communication, computing, and control \cite{Saad2020Vision6G, Wang2023Road6G}. 
	The diverse application scenarios of 6G require not only ultra-reliable and high-capacity communication but also the ability to dynamically shape wireless propagation environments, enabling precise and personalized services for diverse users and scenarios \cite{Wu2020Towards}.
	However, early research based on Shannon's information theorem treated transceivers as fixed and  uncontrollable entities\cite{6773024}, focusing solely on adapting to the electromagnetic environment. These approaches fail to meet the requirements of next-generation communications for shaping the propagation environments.  The advent of multiple-input multiple-output (MIMO) challenges this paradigm via spatial multiplexing and beamforming \cite{1362898,9122596}.
	To realize programmable wireless environments, significant research efforts have focused on the development of flexible antenna technologies, such as reconfigurable intelligent surface (RIS), fluid antenna (FA), and movable antenna (MA), which transformed communication systems from ``adapting" to ``adjusting" the electromagnetic environment \cite{8741198,9264694,10318061}.  A key advantage of these flexible antennas lies in their ability to dynamically reconfigure effective channel gains. Specifically, RIS leverages programmatically controlled reflective components to regulate electromagnetic waves, achieving passive beam shaping and interference mitigation to optimize propagation conditions \cite{8741198}\cite{11072046}.
	FA can employ reconfigurable substances such as liquid metals or ionized solutions with real-time positional adjustments, providing improved flexibility in signal transmission for devices operating in space-limited environments \cite{9264694}.
	MA can adaptively modify their physical locations to change radiation direction and coverage range, enabling responsive adjustments to variations in user positions \cite{10318061}. 
	
	However, the above flexible antenna technologies face critical limitations \cite{yang2025pinching}. RIS suffers significant path loss stemming from dual attenuation, the locations of FA and MA systems are usually fixed or limited within the wavelength scale, which have difficulties in combating large-scale path loss, especially when the LoS link is unavailable \cite{Low-complexity}. Additionally, their limited aperture adjustment range, high manufacturing costs, and  deployment procedures hinder their scalability in complex 6G scenarios \cite{sun_physical_2025}. These challenges have stimulated research into the pinching antenna (PA)  \cite{Suzuki2022Pinching}. Distinguished by its unique pinchable structure, PA can reconstruct large-scale channel and reduce path loss. This design converts path loss into a programmable parameter by enabling flexible adjustment of PA placement along the waveguide, thus changing the distance between radiating points and users. Additionally, this capability allows PA to  establish adjustable LoS links even in complex or obstructed environments without requiring extra hardware, making them a practical solution for dynamically shaping wireless propagation.
	\subsection{Related Works}
	PA can establish LoS links through flexible position adjustment and flexibly scale up or down the system size. These advantages have stimulated research on establishing LoS links through the use of PA to enhance the communication performance in communication systems \cite{PAAC,GPASS}. In \cite{PAAC}, the basic structure and working principle, as well as the signal model in PA-aided systems were introduced.
	The authors in \cite{FlexiblePA} have verified the superior performance of PA compared to the traditional fixed antennas by analyzing the performance in different scenarios.   In \cite{10981775}, the array gain achieved by PA technology was analyzed. The derived closed-form upper bound for array gain shows that it does not always increase monotonically with the number of PA or with decreasing PA spacing. Instead, the optimal number of antennas and spacing must be determined. In \cite{Tyrovolas2025Performance}, the waveguide attenuation was considered to derive  closed-form expressions for outage probability and average rate of PA systems. The above works focused on PA-aided downlink communications. In \cite{10909665,2502.12365}, the uplink communication aided by PAs was investigated. For the first time, the uplink performance optimization of multi-user PA systems was investigated in \cite{10909665}, where it has achieved a balance between throughput and fairness by maximizing the minimum achievable data rate among devices. The authors in \cite{2502.12365} evaluated the performance gains of uplink PA communications systems for three scenarios, i.e., multiple PAs for a single user, a single PA for a single user, and a single PA for multiple users. The results have proved that the flexibility of PAs made the communication systems outperform  conventional fixed antennas-aided communication systems. Beyond the theoretical performance analysis of PA systems, several optimization problems and corresponding algorithms have been formulated to further enhance the performance of such systems \cite{lv_beam_2025, 2502.01590,2502.05917, GPASS}.
	The authors in \cite{lv_beam_2025} designed scalable codebooks and corresponding three-stage  beam training schemes, including non-orthogonal multiple access (NOMA)-aided and dimension-increasing ones for beam training in PA systems under different scenarios, with advantages in reducing training overhead, enhancing flexibility performance, and high-frequency phase alignment. Similarly, issues such as beam training and the joint optimization of antenna position and power have been thoroughly investigated in \cite{2502.01590,2502.05917}. In \cite{GPASS}, the graph neural network was applied to address the joint optimization of antenna placement and power allocation in PA systems. These studies profoundly verify the flexible regulation characteristics of PA systems, whose performance has achieved a significant improvement compared with conventional antenna systems.
	%来点波束的再说人工智能的
	
	Furthermore, due to the outstanding advantages in spectrum efficiency and user access capability, the NOMA technology  has become one of the key technologies for the next-generation communications. The PA-aided NOMA can fully leverage the channel control capability of the former and the multi-user multiplexing advantage of the latter \cite{Low-complexity}\cite{PAAC}\cite{FlexiblePA}\cite{lv_beam_2025} \cite{ActivationPANOMA}\cite{fu_power_2025}. The authors in \cite{FlexiblePA} designed a downlink scheme aided by a PA system that supports both orthogonal multiple access and NOMA, derived the upper bound of system performance and verified the significant performance gains of the  PA-aided NOMA system. They further pointed out that achieving optimal performance requires sophisticated deployment of antenna positions. To address the antenna deployment issue, \cite{Low-complexity} proposed a low-complexity placement design for PAs, aiming to maximize the sum rate of multiple downlink users, and discussed the time division multiple access and NOMA schemes in the scenario of single pinching antenna deployment. 
	The problem of antenna activation in NOMA-aided PA systems was studied in  \cite{ActivationPANOMA}, aiming to maximize the system throughput. Specifically, the study assumed that the power amplifiers had been installed at preconfigured positions prior to transmission, thereby formulating a joint optimization problem involving the number and deployment locations. Beyond the PA-aided NOMA system with a single waveguide, the scenario with multiple waveguides in the NOMA system was studied in \cite{fu_power_2025}. In this research,  a classical power control problem that aims to minimize the total power consumption of all users was formulated. The above research has demonstrated that the two complement each other, forming a high-performance communication solution that provides crucial support for enhancing the 6G communication performance. This integration boasts inherent technical compatibility and substantial research value. %新一代多址技术
	%此外，为契合 6G 对频谱资源高效利用、超高速率及超大容量的核心需求，非正交多址接入（NOMA）技术凭借在频谱效率和用户接入能力方面的突出优势，成为 6G 关键技术之一。夹子天线与 NOMA 技术的结合，能够充分发挥前者的信道调控能力和后者的多用户复用优势，二者相辅相成，形成性能卓越的通信方案，为 6G 通信性能的提升提供重要支撑，这种融合具有天然的技术适配性和极大的研究价值，有望推动 6G 通信技术取得实质性突破。
	
	Due to the openness of PA systems, the security of communication systems is exposed to significant risks. The application of physical layer security in PA systems is an effective way to improve the system's security performance. The authors in \cite{sun_physical_2025} proposed a gradient method and a fractional programming-based block coordinate descent algorithm for single/multi-user wiretap scenarios to optimize baseband beamforming and PA activation positions. The results demonstrate that flexible PA activation serves as an effective PLS measure, enabling PA systems to outperform conventional fixed antenna systems in terms of security.
	In \cite{badarneh_physical-layer_2025}, the authors studied a classic wiretap scenario where confidential information is transmitted from a base station (BS) to a legitimate user, with an eavesdropper attempting interception. The main destination and eavesdropper have random locations. Results show dynamic PA deployment adjustment enhances security performance, and secrecy capacity improves when  PA is placed closer to the destination. 
	%现有的关于物理层安全的论文仅仅局限于波束
	
	%缺少对夹子天线安全性的辅助的NOMA系统中存在用户内部窃听的影响
	%现有的功率模型多数采用等功率模型进行分析，无异于是自断一手臂
	%夹子天线具有极强的灵活度但是多数论文仅仅考虑其在物理空间布置上的灵活度，实际上他还有夹子本身耦合长度的距离。
	%%非常重要的一句话，基站的信号调控，和用户对天线能量的调节，完全是不同的两个维度，后续完全可以发明自调节的天线，进行工厂化的能量部署，定制化服
	%还是得补充一个总的闭式As such, they have to know the messages of the weak users to ensure the successful SIC, which means that NOMA forces the weak users to trust the strong users
	\subsection{Motivation and Contributions}
	Despite progress in PA-aided NOMA  systems and PA-based PLS, critical gaps remain. In particular, the use of successive interference cancellation (SIC) leads to an internal eavesdropping of NOMA with multi-user, where a public information user may overhear the information of confidential information users in the superimposed information streams. However, no relevant research on PA-aided NOMA systems against internal eavesdropping has been reported in the existing literature yet. Existing works, such as \cite{Low-complexity}\cite{PAAC}\cite{FlexiblePA}\cite{lv_beam_2025} \cite{ActivationPANOMA}\cite{fu_power_2025}, mainly focus on improving the robust performance of  PA-aided  NOMA systems but overlook the investigation of internal security issues. %Furthermore, PLS research in PA systems  is mainly limited to multi-waveguide systems, lacking analysis of single-waveguide deployments, a practical low-cost scenario 
	In the research on PLS for PA systems, the single-waveguide multi-PA systems primarily focus on adjusting the phase of PAs to enhance system security, while the multi-waveguide and multi-PA systems concentrates on optimizing secure beamforming schemes \cite{sun_physical_2025,badarneh_physical-layer_2025}.  Current literature universally identifies flexibility as the most significant characteristic of PAs. However, such flexibility is limited to the flexible addition/removal and location deployment of PAs. The two common power models, i.e., the equal power model and the proportional power model fix the radiated power of PAs, which limits the flexibility of their radiated power.
	These gaps hinder the further development and application of PA-aided NOMA  systems in  practical secure scenarios. Motivated by the above, we analyze the security performance of  PA-aided NOMA systems with internal eavesdropper and propose a flexible power strategy to enhance the system security. The main contributions of this paper are summarized as follows:
	\begin{itemize}
		\item To the best of the authors, we first fill the internal eavesdropping  research gap in existing studies on PA systems. In this paper, we consider a typical NOMA system with two users, where a public information user close to the BS has a potential risk of intercepting the confidential information of  user far away from the BS. In the conventional fixed antenna system, due to a better channel condition of the public information user, the confidential information is decoded by the public information user with a high probability. However, PA systems can enhance the channel condition of the far confidential information user while weakening that of the near public user. When  public information user's channel condition is no worse than that of the confidential information user, PA systems can  enhance the system security by differentiated power allocation, i.e., more power to the PA serving the confidential information user and less power to that for the public information user. 
		\item  We analyze the SOP for the PA-aided NOMA systems with internal eavesdropping. We derive exact closed-form expressions for SOP and perform asymptotic analysis to reveal the impact of PA in enhancing the secrecy performance.   Since the two conventional power models, i.e., the equal power model and the proportional power model,  allocate less or the same power to the PA farther from the BS, they cannot effectively address scenarios with severe security risks.  To address this issue, we propose a flexible power strategy, which involves adjusting the radiated power by using a PA  with different coupling lengths or altering the coupling lengths between the PAs and the waveguide. We formulate an optimization problem for improving system security by optimizing the coupling length to regulate the radiated power, and the conclusions derived from the solution process can the PA-aided NOMA system to achieve minimum  SOP.
		% By optimizing the PA coupling length to regulate radiated power, the system security is improved. The proposed optimization problem can effectively obtain the minimum value of SOP.
		\item The results shows that: 1) Compared to conventional fixed antennas, in adverse security scenarios where internal eavesdropping is near BS,   PA  can significantly enhance system security by adjusting the PA coupling length; 2) In the NOMA system, to guarantee communication of both public information user and confidential information user while preventing confidential information leakage to public information users, an effective strategy is to set  the signal power allocation coefficient of confidential information user much lower than that of public information user. On the premise of ensuring no communication outage for public information user, more power is allocated to confidential information user; 3) We define the maximum eavesdropping distance and maximum reliable transmission distance for user activities. As the activity cell of public information user expands, the SOP decreases; whereas as the activity cell of confidential information user expands, the SOP increases.
	\end{itemize}
	\subsection{Organization}
	The remainder of this paper is organized as follows. Section \ref{section 2} introduces a two PAs-aided  secure NOMA system with a single waveguide and two paired users, i.e.,  a far confidential information user and a near public information user attempting to eavesdrop on the confidential information. Section \ref{section 3} analyzes the performance of the NOMA system achieved by PA, and derives the exact and asymptotic closed-form expressions of SOP to gain useful insights. In Section \ref{section optimal problem}, we propose the flexible power strategy to improve the holistic performance of the system. To guide users in adjusting the antenna coupling length in different scenarios for achieving a secure  transmission,  an optimization problem based on proposed strategy is formulated. In Section \ref{section 4}, the numerical results are presented to verify the accuracy of theoretical analysis and  the effectiveness of the proposed strategy.
	Finally, Section \ref{section 5} concludes the paper and summarizes the key findings.
	\section{System Model}\label{section 2}
	\subsection{System Topology}
\begin{figure}[t]
		\centering
		\includegraphics[width=3.5in]{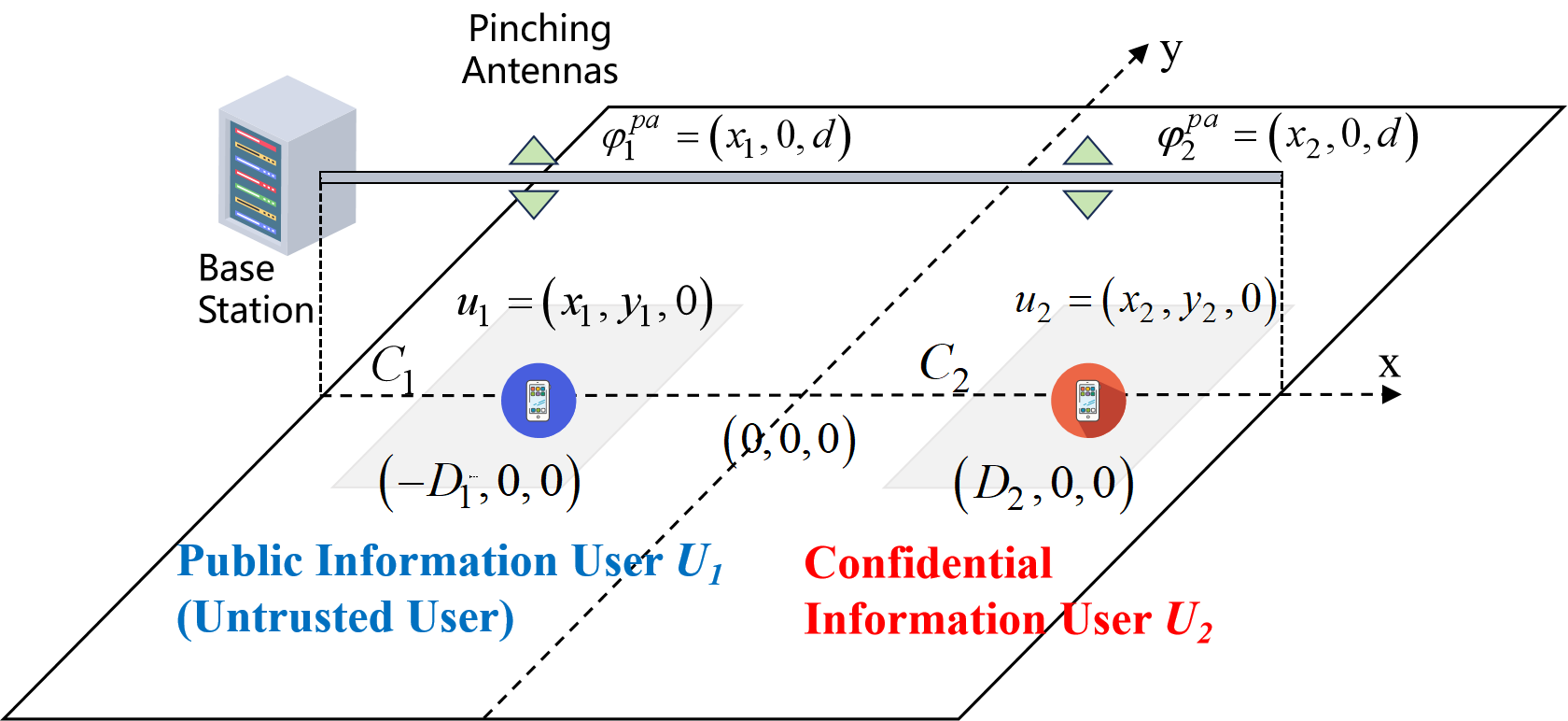}
		\caption{The two pinching antennas-aided NOMA system, with a single waveguide,  a near public information user, and a far confidential information user.}
		\label{fig:systemmodel}
	\end{figure}

	As shown in Fig. \ref{fig:systemmodel}, we consider a PA-aided NOMA system consisting of a BS,  two NOMA users ($U_{1}$ and $U_{2}$), a  dielectric waveguide\footnote{Compare to free-space propagation, the power attenuates in waveguide
		verges on being negligible, e.g., approximately 0.01-0.03 dB/m for a circular
		copper waveguide at 15 GHz, and 0.1 dB/m at 28 GHz \cite{Pozar2011}. This study
		aims to systematically analyze the potential security vulnerabilities of internal
		eavesdropping in PA-assisted NOMA systems. In follow-up investigations,
		the influence of waveguide losses on security performance will be integrally
		incorporated into the analytical framework.}, and two PAs (PA-1 and PA-2). We consider the case that the number of user is equal to that of PA as the users are far apart from each other. Specifically, $U_{1}$ is closer to the BS, while $U_{2}$ is farther from the BS. Assume that the $i$-th PA is placed closest to $U_{i}$, where $i\in\{1,2\}$ Among the users, $U_{1}$ is a public information user and $U_{2}$ is a confidential information user. Both $U_1$ and $U_2$ receive the information in accordance with the NOMA protocol and act as regular transceivers in the communication network. 	Each user is equipped with a single antenna. In practical application scenarios, there is a clear distinction between the information requirements of two types of users.
	Users with a demand for confidential information transmission typically focus on scenarios involving the transmission of personal private information (such as identity privacy, and  private conversations) or sensitive account information (such as payment passwords and  account permission data).
	By contrast, users who need to receive public information mainly correspond to scenarios of accessing public information that does not involve sensitive data, such as browsing news and watching entertainment videos. 
	
	The waveguide is oriented along the $x$-axis at a height of $d$. In this paper, we consider a hostile eavesdropping case, where the untrusted user $U_{1}$ is closer to the BS, while the confidential information user $U_{2}$ is farther from the BS. The untrusted user can be either the near or the far, but we focus on this case because it is in general more challenging\footnote{The conventional power models of PA, namely the equal power model and the proportional power model, allocate more or an equal amount of  power to $U_{1}$, which increase the risk of confidential information leakage. Thus, the scenario where the eavesdropper is closer to the BS is  more detrimental compared with that where the eavesdropper is far from the BS. This conclusion is further illustrated in the Sections \ref{section 3}, \ref{section optimal problem} and \ref{section 4}.}. Since multiple users receive the superimposed signal at the same time, confidential information faces a potential risk of being leaked to \(U_{1}\). Specifically, $U_1$ will employ the successive interference cancellation (SIC) technique to decode $U_2$’s confidential information from the received superimposed signal after successfully decoding its own information. As such, $U_1$ poses a potential  eavesdropping threat to $U_2$ as an internal eavesdropper of NOMA.
	Unlike conventional external eavesdroppers, $U_1$ has a dual identity in the system. On one hand, as a NOMA user operating in accordance with the intended protocol, $U_1$’s transmission performance needs to be ensured. On the other hand, as an internal user with potential eavesdropping risks, the security issue needs to be  tackled to prevent $U_1$ from successfully intercepting the confidential information of $U_{2}$.
	
	The spatial coordinates of PAs are denoted as $\varphi_{i}^{pa} = \left(x_{i}, 0, d\right)$, while the positions of the two NOMA users are specified by $u_{i} = \left(x_{i}, y_{i}, 0\right)$%\footnote{The flexibility and portability of PAs enable them to be deployed in close proximity to users, thereby mitigating critical electromagnetic attenuation (including shadow fading and free-space path loss).}
	.
	The inherent characteristics of PAs allow for dynamic positional adjustment in response to changes in user positions.
	Specifically, it is assumed that the two NOMA users  are uniformly distributed in two square cells, i.e., $C_{1}$ with center at $\left(-D_{1}, 0, 0\right)$ and $C_2$ with  center at $\left(D_{2}, 0, 0\right)$, respectively. Both side lengths of  $C_{1}$ and $C_{2}$ are equal to $D$.   
	Owing to the substantial inter-user distance, the signal strength received by either user from the PA serving the other user is extremely attenuated. Consequently, the deployment of a single PA is insufficient to fulfill the transmission requirements of NOMA. Therefore, each user is  paired with one dedicated PA to meet the operational demands of NOMA communications. 
	\subsection{Signal Model}
	In the PAs aided NOMA system, the signal received  at $U_i$ can be expressed as
	\begin{align}
		y_{i} = \mathbf{h}_{i}^{H}\mathbf{s} + 	w_{i},
		\label{received signal}
	\end{align}
	where $w_{i}$ denotes additive Gaussian noise with zero mean  and variance $\sigma^{2}$.  The channel vector can be expressed as  \cite{zhang2022beam, FlexiblePA}
	\begin{align}
		\mathbf{h}_{i} = \left[\frac{\eta^{\frac{1}{2}}  e^{-j\frac{2\pi}{\lambda}\Vert{u_i}-{\varphi^{pa}_{1}}\Vert} }{\Vert{u_i}-{\varphi^{pa}_{1}}\Vert} e^{-j\theta_{1}},\frac{\eta^{\frac{1}{2}}  e^{-j\frac{2\pi}{\lambda}\Vert{u_i}-{\varphi^{pa}_{2}}\Vert} }{\Vert{u_i}-{\varphi^{pa}_{2}}\Vert}  e^{-j\theta_{2}}
		\right]^{T},
		\label{channel vector}
	\end{align}
	where $\eta = \frac{\lambda^2}{16\pi^2}$ represents the complex channel of the free-space propagation at the reference distance of 1 meter, $\lambda = \frac{c}{f_{c}}$ denotes the free-space wave length, $c$ is the speed of light, $f_{c}$ is the carrier frequency, $j$ is the imaginary unit, $\Vert \cdot \Vert$ denotes the Euclidean norm, $\theta_{i} = \frac{2\pi}{\lambda_g}\Vert\varphi_{bs} - \varphi^{pa}_{i} \Vert$ is the phase shift experienced at the $i$-th PA, $\lambda_g = \lambda/n_{eff}$ is the waveguide wavelength in the dielectric waveguide, $n_{eff}$ is the effective refractive index of a dielectric waveguide, and $\varphi_{bs} = \left(x_{0}, 0, d\right)$ denotes the position of BS.
	
	The two PAs deployed on a single waveguide serve $U_{1}$ and $U_{2}$ via NOMA technology, and the the superimposed signal  $\mathbf{s}$ in  \eqref{received signal} for the users can be expressed as \cite{ActivationPANOMA,7676258}
	\begin{align}
		\mathbf{s} = \left[\sqrt{p_{1}},\sqrt{p_{2}}\right]^{T}\left(\sqrt{\alpha_{1}}s_{1}+\sqrt{\alpha_{2}}s_{2}\right),
		\label{s vector}
	\end{align}
	where $p_{i} = P\cdot \epsilon_{i} $ denotes the transmission power at the $i$-th PA, $P$ is the transmission power of BS, $\epsilon_{i}$ is the normalized power coefficient of the $i$-th PA, $s= \left(\sqrt{\alpha_{1}}s_{1} + \sqrt{\alpha_{2}}s_{2} \right)$ is the superimposed signal,  $s_{i}$ denotes $U_{i}$'s signal, $\alpha_{i}$ denotes the power allocation coefficient for $U_{i}$, and $\alpha_{1}  +  \alpha_{2} = 1$. 
	According to the coupled-mode theory, $\epsilon_{i}$ can be expressed as  \cite{7654321}
	\begin{align}\label{power coefficient}
		\epsilon_{i} = 
		\begin{cases}
			F\sin^{2}\left(\kappa L_{1}\right)&, i=1, \\
			%\left(1-\sum\limits_{j=1}^{i-1}\epsilon_{j}\right)F\sin^{2}\left(\kappa L_{i}\right)&, i\ge 2,
			\left(1-F\sin^{2}\left(\kappa L_{1}\right)\right)F\sin^{2}\left(\kappa L_{2}\right)&, i= 2,
		\end{cases} 
	\end{align}
	where $ 0 < F \le 1$ is the coupling efficiency, $\kappa$ is the coupling coefficient, and  $L_{i}$ is the coupling length of the $i$-th PA. Substituting \eqref{channel vector}  and \eqref{s vector} into \eqref{received signal}, the received signal at $U_{i}$ can be expressed as 
	\begin{align}
		y_{i} =\sum_{n=1}^{2}p_{i}\frac{\eta^{\frac{1}{2}}  e^{-j\frac{2\pi}{\lambda}\Vert{u_i}-{\varphi^{pa}_{n}}\Vert} }{\Vert{u_i}-{\varphi^{pa}_{n}}\Vert} e^{-j\theta_{n}}s + w_{i}.
	\end{align}
	
	The channel strength of $U_{i}$ is denote by $|h_{i}|^{2}$, where $h_{i} =\sum_{n=1}^{2}\frac{\eta^{\frac{1}{2}}  e^{-j\frac{2\pi}{\lambda}\Vert{u_i}-{\varphi^{pa}_{n}}\Vert} }{\Vert{u_i}-{\varphi^{pa}_{n}}\Vert} e^{-j\theta_{n}}$. Since the two users and their corresponding active cells are far apart, large-scale path loss dominates the channel gain, and the distance between \(U_i\) and the PAs becomes the key factor. Moving a PA a few wavelengths to satisfy $\frac{2\pi}{\lambda}\Vert u_{i} - \varphi_{i}^{pa}  \Vert = 2k\pi$  has a limited impact on the distance between PA and user, where $k$ is an arbitrary integer. Therefore, $h_{i}$ can be simplified as \cite{Low-complexity}
	\begin{align}
		\vert h_{i} \vert^{2} = \frac{\eta}{\Vert u_{i} - \varphi_{i}^{pa} \ \Vert^{2}}.
		\label{simple channel}
	\end{align}
	Furthermore, given that $D <\vert D_{1} + D_{2}\vert$, this condition ensures the two activity cells are separated from each other and have no overlapping areas. %Though the distance between $\varphi_{1}^{pa}$ and BS is closer than that between $\varphi_{2}^{pa}$ and BS, the distance relationship between $i$-th PA and $i$-th user determines that $\vert h_{1} \vert^2 $ and $\vert h_{2} \vert^2$ are similar.
	In the equal power model, the two PAs deployed on a single  waveguide have the same radiated power. Based on this characteristic, the  PA-aided NOMA system can be equivalently regarded as two conventional antennas with consistent radiation intensity, and these two antennas are arranged in parallel along the x-axis. This system design enables the establishment of a robust communication link, but it also increases the risk of confidential signals leakage. 
	
	To ensure secure transmission, the signal power coefficients are optimized in line with the priority-based NOMA power allocation principle. Specifically, a higher signal power coefficient is assigned to the public information user (prone to eavesdropping), while a lower one is allocated to the user with high security clearance, i.e., \(\alpha_1 > \alpha_2\).
	The core of the SIC technique lies in the sequential decoding process of multi-user signals. First, the system prioritizes and sorts multi-user signals. Then, when a user decodes the target signal, it treats the signals of other users as interference. As such, $U_1$ treats $U_2$’s signal as interference and directly decodes its own information. While a less power is allocated to $U_{2}$, and $U_{2}$ first decodes $U_{1}$'s information successfully. Then, $U_{2}$ eliminates the interference from $U_1$’s signal based on this decoded information, and then decodes its own information. A critical prerequisite for this process is that $U_2$ obtains $U_1$’s information to ensure the success of SIC.
	In this process, the public information user with potential eavesdropping risk attempts to decode the confidential information signal $s_{2}$ from the received superimposed signal.  The signal-to-interference-plus-noise ratio (SINR) for $U_i$ to decode information signal ($s_{1}$ or $s_{2}$) is denoted by $\gamma_{u_{i},j}$, where $i\in\{1,2\}$ denotes $U_{1}$ and $U_{2}$, and $j\in \{1,2\}$ denotes signal $s_{1}$ and $s_{2}$.  Based on the above analysis, the SINRs can be expressed as
	\begin{align}
		\gamma_{u_1,1} &= \frac{\alpha_{1} \rho_1 \vert h_1\vert^2}{\alpha_2 \rho_{1} \vert h_1 \vert^2 + 1}, \label{u1,1}\\
		\gamma_{u_1,2} &= \alpha_2\rho_{1} \vert h_1 \vert^2 ,\label{u1,2}\\
		\gamma_{u_2,1} &= \frac{\alpha_1\rho_{2} \vert h_{2} \vert^2}{\alpha_2\rho_{2} \vert h_2 \vert^{2}+1} ,\label{u2,1}\\
		\gamma_{u_2,2} &= \alpha_2 \rho_{2} \vert h_2 \label{u2,2}\vert^2,
	\end{align}
	where $\rho_{i}  = \rho_{t}\epsilon_{i}$ denotes the transmit signal-to-noise ratio (SNR) at the $i$-th PA, and  $\rho_{t} = \frac{P}{\sigma^{2}}$ denotes the SNR at the BS.
	\section{Performance Analysis} \label{section 3}
	In this section, we analyze the secrecy performance of the NOMA system achieved by PAs and derive the closed-form SOP. Further, asymptotic analysis is performed to gain deep  insights. In particular, a flexible power  strategy is proposed by building upon two conventional power models of PAs. Specifically, by regulating the coupling lengths of PAs, a novel degree of freedom in power allocation is introduced into the   PA-aided NOMA system, thereby effectively enhancing the system security. 
	
	We define the secrecy outage event of the PA-aided NOMA system as the following three scenarios, where $\gamma_{i}$ denotes the minimum decoding SINR threshold of signal $s_{i}$. 
	\begin{itemize}
		\item$U_{1}$ fails to decode the confidential signal $s_{1}$, i.e., $\left\{\gamma_{u_1,1} <\gamma_{1} \right\}$.	 
		\item$U_{1}$ succeeds in eavesdropping on $s_{2}$ i.e, $\left\{\gamma_{u_1,2}\ge\gamma_{2}\right\}$. 
		\item$U_{2}$ fails to decode its own signal $s_{2}$, i.e., $\left\{\gamma_{u_2,1} < \gamma_{1}\cup\gamma_{u2,2} < \gamma_{2}\right\}$.
	\end{itemize}
	Compared with the SOPs in \cite{10915642,8647889}, the SOP defined in this work considers two key aspects, i.e., the reliability that both users satisfy the minimum decoding threshold, and the secure transmission requirement of preventing the internal eavesdropper $U_1$ from decoding $s_2$. Therefore, the SOP can be expressed as
	\begin{align}
		\text{P}_{\text{sop}} =& \text{ Pr}\left[\gamma_{u_1,1} < \gamma_{1}   \cup\gamma_{u_1,2} \ge \gamma_{2}   \cup\gamma_{u_2,1} < \gamma_{1}   \cup\gamma_{u_2,2} < \gamma_{2}   \right]  	\notag
		\\ 
		=&1-\text{ Pr}\left[\gamma_{u_1,1} \ge \gamma_{1},\gamma_{u_1,2} < \gamma_{2}, \gamma_{u_2,1} \ge\gamma_1 ,\gamma_{u_2,2} \ge \gamma_{2}\right].
		\label{define sop}
	\end{align}

	Substituting \eqref{simple channel}–\eqref{u2,1} into \eqref{define sop}, the SOP is derived as
	\begin{align}
		\text{P}_{\text{sop}} =& 1 - \text{Pr}\Bigg[\frac{\eta \alpha_1\rho_{1} / \left(y_{1}^{2} + d^2\right)}{\eta\alpha_2\rho_{1}/\left(y_{1}^{2}+d^2\right)+1} \ge \gamma_{1}, \frac{\eta \alpha_2\rho_{1}}{y_1^2+d^2} < \gamma_{2}, \notag \\   
		&  \frac{\eta \alpha_1\rho_{2} / \left(y_{2}^{2} + d^2\right)}{\eta\alpha_2\rho_{2}/\left(y_{2}^{2}+d^2\right)+1}  \ge \gamma_{1}, 	 \frac{\eta \alpha_2\rho_{2}}{y_2^2+d^2}   \ge \gamma_{2}
		\Bigg] \notag\\
		=&1-\text{Pr}\Bigg[\frac{\eta\alpha_{2}\rho_{1}}{\gamma_{2}}-d^{2} < y_{1}^{2} \le \frac{\eta\alpha_{1}\rho_{1}}{\gamma_{1}}-\eta\alpha_{2}\rho_{1} -d^{2},	\notag\\
		&    y_2^2\le \frac{\eta \alpha_1 \rho_{2}}{\gamma_{1}} - \eta \alpha_2\rho_{2} - d^2, y_2^2 \le\frac{\eta \alpha_2 \rho_{2} }{ \gamma_2}-d^2  \Bigg]\notag\\
		=& 1- \text{ Pr}\left[\omega_{1} <y_{1}^{2}\le\omega_{2}, y_{2}^{2}\le \min\left(\omega_{3},\omega_{4}\right)
		\right] \notag \\
		=&  1-
		\underbrace{ \text{ Pr}\left[ \omega_{1} <y_{1}^{2}\le\omega_{2}	\right] }_{ \Omega_{1}}
		\underbrace{  \text{Pr}\left[y_{2}^{2}\le\min\left(\omega_{3},\omega_{4}\right) \right] }_{\Omega_{2}}		 ,
		%符号问题后面再改
		\label{y sop}
	\end{align}
	where $ \omega_{1}= \frac{\eta \alpha_{2} \rho_{1} }{\gamma_{2}}-d^{2}$, 
	$\omega_{2}= \frac{\eta \alpha_{1} \rho_{1} }{\gamma_{1}}- \eta \alpha_{2}\rho_{1}-d^{2}$, 
	$ \omega_{3}= \frac{\eta \alpha_{1} \rho_{2} }{\gamma_{1}}-\eta\alpha_{2}\rho_{2} - d^{2}$, 
	$\omega_{4}= \frac{\eta \alpha_{2} \rho_{2} }{\gamma_{2}}-d^{2}$, $\Omega_{1}$ denotes the probability that $U_{1}$ achieves reliable transmission and the information of $U_{2}$ is not leaked, and $\Omega_{1}$ denotes the probability that $U_{2}$ achieves reliable transmission. 
	Both $y_{1}$ and $y_{2}$ follow a uniform distribution, and their probability density functions (PDFs) are given respectively as 
	\begin{align}
		f_{Y_{1}}(y_{1}) =& 
		\begin{cases} 
			\frac{1}{D}, & \text{for } -\frac{D}{2} \leq y_1 \leq \frac{D}{2}, \\
			0, & \text{otherwise}.
		\end{cases} \label{pdf y1}\\
		f_{Y_2}(y_2) =& 
		\begin{cases}
			\frac{1}{D}, &\text{for } -\frac{D}{2} \leq y_2 \leq \frac{D}{2}, \\
			0, & \text{otherwise}.\label{pdf y2}
		\end{cases}
	\end{align}
	Using \eqref{pdf y1} and \eqref{pdf y2}, the SOP for PAs-aided NOMA systems is presented in the following Lemmas and theorems.
	\begin{lem}\label{PPP1 }
		The exact closed-form expression for	$\Omega_{1}$ in \eqref{y sop} can be derived as 
		\begin{align}\label{varOmega1}
			\Omega_{1} = 
			\begin{cases}
				1, & \omega_1 \leq 0, \omega_2 \geq \frac{D^2}{4}, \\
				\frac{2\sqrt{\omega_2}}{D}, & \omega_1 \leq 0, 0 < \omega_2 < \frac{D^2}{4}, \\
				\frac{2\left(\sqrt{\omega_2} - \sqrt{\omega_1}\right)}{D}, & 0 < \omega_1 < \omega_2 < 	\frac{D^2}{4}, \\
				\frac{2\left(\frac{D}{2} - \sqrt{\omega_1}\right)}{D}, & 0 < \omega_1 < \frac{D^2}{4}, \omega_2 \geq \frac{D^2}{4}, \\
				0, & \text{otherwise}.
			\end{cases}
		\end{align}
		\begin{IEEEproof}
			See Appendix \ref{ProfL1}.
		\end{IEEEproof}
		\begin{rem}
			Since $\Omega_{1}$  represents the probability of ensuring $U_{1}$'s reliability while preventing $U_{2}$'s information from being intercepted by $U_{1}$, we define $\sqrt{\omega_{1}}$ as the maximum distance at which  $U_{1}$ can successfully eavesdrop. When the projection of $U_{1}$'s distance from the center of the $C_{1}$ on the y-axis is longer than $\sqrt{\omega_{1}}$, $U_{1}$ cannot decode the confidential information of $U_{2}$. We define $\sqrt{\omega_{2}}$  as the maximum reliable transmission distance for $U_{1}$. The signal $s_{1}$ can be decoded by $U_{1}$, when  the projection of $U_1$'s distance from the center of the active area on the y-axis is shorter than $\sqrt{\omega_2}$. 
		\end{rem}\label{rem0}
		
		%	\begin{rem}
			%	From  the closed-form expression of $\Omega_{1}$ in \eqref{varOmega1}, it can be  observed that $\omega_{1}$ is closely associated with the probability of confidential information leakage. As the decoding threshold \(\gamma_{2}\) increases, \(\omega_{1}\) decreases, and the security performance deteriorates. When \(\omega_{1}\) increases, the coverage range of \(y_{1}\) decreases, which leads to occurrences of security outage events with a high probability. The increase of  \(\gamma_{2}\) decreases threshold for \(U_{1}\) to decode confidential information becomes. This reduces the difficulty for \(U_{1}\) to obtain the confidential information, thereby undermining security. A decrease in \(\gamma_{1}\) reduces the difficulty of decoding \(s_{1}\), which in turn increases \(\omega_{2}\). The increase of $\Omega_{1}$ improves the reliability performance of $U_{1}$ and the security  performance of $U_{2}$.
			%	\end{rem}\label{rm 1}
	\end{lem}
	\begin{lem}\label{PPP2}
		The expression of $\Omega_{2}$ in \eqref{y sop} is shown as
		\begin{align} \label{Omega2}
			\Omega_{2} = 
			\begin{cases}
				1, &\min\left(\omega_{3},\omega_{4} \right) \ge \frac{D^{2}}{4},\\
				\frac{2\sqrt{\min\left(\omega_{3},\omega_{4}\right)}}{D}, &0<\min\left(\omega_{3},\omega_{4} \right)<\frac{D^{2}}{4},\\
				0, &\min\left(\omega_{3},\omega_{4}\right)\le 0.
			\end{cases}
		\end{align}
		\begin{IEEEproof}
			%The feasible region is $0\le\min(\omega_{3},\omega_{4})\le\frac{D^{2}}{4}$.	
			$U_{2}$ achieves reliable transmission, i.e., $\Omega_{2} = 1$,	when the dimensions of $U_{2}$'s activity cell is shorter than  $\sqrt{\min(\omega_{3},\omega_{4})}$.  The proof of $\Omega_{2}$ is given by 
			\begin{align}\label{pr O2}
				\Omega_{2} =& \text{Pr}\left[y_{2}^{2}\le\min\left(\omega_{3},\omega_{4}\right)\right]\notag \\
				=
				&\frac{1}{D}\Big[
				\mathbb{I}\left(\min(\omega_{3},\omega_{4})\ge\frac{D^{2}}{4}\right)\int_{-\frac{D}{2}}^{-\frac{D}{2}}dy_{2}\notag \\
				&+ \mathbb{I}\left(0<\min(\omega_{3},\omega_{4})<\frac{D^2}{4}\right)\int_{-\sqrt{\min(\omega_{3},\omega_{4})}}^{\sqrt{\min(\omega_{3},\omega_{4})}}dy_{2}
				\Big],
			\end{align}
			where $\mathbb{I}\left(X\right)$ denotes the indicator function, which is defined as 
			\begin{align}
				\mathbb{I}\left(X\right) = 
				\begin{cases}
					1, & \text{if X is true},\\
					0, &\text{otherwise}.
				\end{cases}
			\end{align}
			Through mathematical manipulations on \eqref{pr O2}, we derive the closed-form expressions for $\Omega_{2}$ as presented in \eqref{Omega2}, thereby finalizing the proof.
		\end{IEEEproof}
		\begin{rem}
			From Lemma \ref{PPP2}, it can be observed that \eqref{Omega2} characterizes the robustness of $U_2$'s information transmission. Similarly defined to $\sqrt{\omega_{2}}$, $\sqrt{\min\left(\omega_{3},\omega_{4}\right)}$ is defined as the maximum reliable transmission distance for $U_{2}$.  When $U_i$ is positioned at the geometric center of $C_i$, the minimal distance between PAs and users results in enhanced channel. As $\alpha_{2}$ is set to be small to ensure $U_{1}$ cannot decode $s_2$,   the system is required to   guarantee the successful  decoding of $ s_1 $ by $U_{1}$ and $ s_2 $ by $U_{2}$. Consequently, an effective solution is to allocate more power to the far user by adjusting the coupling lengths 
		\end{rem}\label{rm 2}
	\end{lem}
	\begin{thm}\label{thm3}
		With the helps of Lemmas \ref{PPP1 } and \ref{PPP2}, the SOP is given by \eqref{eq:SOP_condensed}, which is shown at the top of the next page.
	\begin{IEEEproof}
	The piecewise expressions of \(\Omega_1\) and \(\Omega_2\) are derived in Lemmas \ref{PPP1 } and \ref{PPP2}, respectively, with their segmentations determined by the relative magnitudes of \(\omega_1, \omega_2, \min(\omega_3,\omega_4)\), and \(\frac{D^2}{4}\). The product \(\Omega_1 \Omega_2\) is obtained by  combining all  piecewise conditions of \(\Omega_1\) and \(\Omega_2\), followed by substituting their corresponding segment expressions and performing simple algebraic multiplication for each combination. Given the definition \(\text{P}_{\text{sop}} = 1 - \Omega_1 \Omega_2\), the closed-form expression of \(\text{P}_{\text{sop}}\)  in \eqref{eq:SOP_condensed} is thereby obtained. Thus, the proof is completed.
	\end{IEEEproof}
	\end{thm}
	\begin{figure*}[t!]
		\begin{align}\label{eq:SOP_condensed}
			\text{P}_{\text{sop}} = 1 -
			\begin{cases}
				1, 
				&\omega_{1} \le 0, \omega_{2}\ge\frac{D^{2}}{4}, \min(\omega_{3},\omega_{4})\ge\frac{D^{2}}{4},\\
				\frac{2\sqrt{\omega_{2}}}{D}, 
				&\omega_{1}\le0, 0<\omega_{2}<\frac{D^{2}}{4}, \min(\omega_{3},\omega_{4})\ge\frac{D^{2}}{4},\\
				\frac{2\left(\sqrt{\omega_{2}}-\sqrt{\omega_{1}}\right)}{D}, 
				&0<\omega_{1}<\omega_{2}<\frac{D^{2}}{4}, \min(\omega_{3},\omega_{4})\ge\frac{D^{2}}{4},\\
				\frac{2\left(\frac{D}{2}-\sqrt{\omega_{1}}\right)}{D}, 
				&0<\omega_{1}<\frac{D^{2}}{4}, \omega_{2}>\frac{D^{2}}{4}, \min(\omega_{3},\omega_{4})\ge\frac{D^{2}}{4},\\
				\frac{2\sqrt{\min\left(\omega_{3},\omega_{4}\right)}}{D},
				&\omega_{1}\le0, \omega_{2}\ge\frac{D^{2}}{4}, 0 < \min\left(\omega_{3},\omega_{4}\right) <\frac{D^{2}}{4},\\
				\frac{4\sqrt{\omega_{2}}
					\sqrt{\min\left(\omega_{3},\omega_{4}\right)}}{D^{2}},
				&\omega_{1}\le0, 0<\omega_{2}<\frac{D^{2}}{4}, 0 < \min\left(\omega_{3},\omega_{4}\right) <\frac{D^{2}}{4},\\
				\frac{4\left(\sqrt{\omega_{2}}- \sqrt{\omega_{1}}\right)
					\sqrt{\min\left(\omega_{3},\omega_{4}\right)}}{D^{2}},
				&0<\omega_{1}<\omega_{2}<\frac{D^{2}}{4}, 0 < \min\left(\omega_{3},\omega_{4}\right) <\frac{D^{2}}{4},\\
				\frac{4\left(\frac{D}{2}-\sqrt{\omega_{1}}\right)
					\sqrt{\min\left(\omega_{3},\omega_{4}\right)}}{D^{2}}, 
				&0<\omega_{1}<\frac{D^{2}}{4}, \omega_{2}>\frac{D^{2}}{4}, 0 < \min\left(\omega_{3},\omega_{4}\right) <\frac{D^{2}}{4},\\
				0, &\text{otherwise}.
			\end{cases}
		\end{align}
		\hrulefill
	\end{figure*}  
	\begin{rem}\label{rem3}
		From Theorem \ref{thm3}, 
		ensuring the reliable transmission of $U_{1}$ and $U_{2}$ while preventing the leakage of  confidential information is the key challenge in the PA-aided NOMA systems. Power allocation is required  to satisfy the requirements of all authorized users. At the signal power allocation level, the ratio of \(\alpha_{1}\) to \(\alpha_{2}\) needs to exceed the user’s decoding threshold, while less power should be allocated to confidential signal \(s_{2}\) to enhance its security. To further address these issues, at the PA radiated power level, the transmit SNR at PA-1 (\(\rho_1\)) can be reduced (while meeting \(U_{1}\)'s decoding threshold) to make \(\gamma_{u1,2} < \gamma_{2}\), and the transmit SNR at PA-2 (\(\rho_{2}\)) can be increased to improve the reliability of \(U_2\). 
		In conventional fixed antenna-aided NOMA communication systems, when an internal eavesdropper is close to the BS, its channel condition is significantly favorable. This results in NOMA systems with a high risk of confidential information leakage.
		PAs address this by regulating radiated power through antenna design, enhancing system security at minimal cost.
		%再下一章节说明传统两种模式的弊端。
	\end{rem}
	To obtain further insights, we examine the system's asymptotic characteristics under high transmission SINR, i.e., $\rho_{i}\to \infty$. 
	\begin{cor}
		The asymptotic analysis of SOP is given by
		\begin{align}
			\lim_{\rho_{t}\to \infty} \text{P}_{\text{sop}} = 1.
		\end{align}
		
		\begin{IEEEproof}
			As \(\rho_{t}\to\infty\), both \(\omega_{1}\) and \(\omega_{4}\) approach  infinity. Combining this with \eqref{varOmega1}, it follows that \(\Omega_{1} = 0\). Further, from \eqref{Omega2}, it is observed that as \(\rho_{t}\to\infty\), the asymptotic behavior of \(\omega_{3}\) and the corresponding value of \(\Omega_{2}\) are determined by the sign of \(\eta\left(\frac{\alpha_{1}}{\gamma_{1}} - \alpha_{2}\right)\). Specifically, if \(\eta\left(\frac{\alpha_{1}}{\gamma_{1}} - \alpha_{2}\right) > 0\), \(\omega_{3}\) approaches \(+\infty\) and \(\Omega_{2} = 1\); if \(\eta\left(\frac{\alpha_{1}}{\gamma_{1}} - \alpha_{2}\right) < 0\), \(\omega_{3}\) approaches \(-\infty\) and \(\Omega_{2} = 0\); and if \(\eta\left(\frac{\alpha_{1}}{\gamma_{1}} - \alpha_{2}\right) = 0\), \(\omega_{3} = -d^2\) with \(\Omega_{2} = 0\). Substituting the aforementioned values of \(\Omega_{1}\) and \(\Omega_{2}\) into \eqref{y sop}  completes the proof.
		\end{IEEEproof}
	\end{cor}\label{cor 1} 
	
	%	\begin{cor}
		%		Asymptotic analysis of \eqref{u1,2}-\eqref{u2,1} reveals:
		%		\begin{equation}
			%			\lim_{\rho_{1} \to \infty}\gamma_{u_1,1} \approx 	\frac{\alpha_1}{\alpha_2} \quad \text{and} \quad \lim_{\rho_{1} \to \infty}\gamma_{u_1,2} \approx +\infty.
			%			\label{eq:asymptotic_sinr}
			%		\end{equation}
		%		\begin{equation}
			%			\lim_{\rho_{2}\to \infty}\gamma_{u_2,1} \approx \frac{\alpha_{1}}{\alpha_{2}} \quad \text{and} \quad \lim_{\rho_{2} \to\infty}\gamma_{u_2,2} \approx +\infty.
			%			\label{asymptotic2}
			%		\end{equation}
		%		As the transmit SNR $\rho_{i} \to \infty $, the SINR for decoding $s_{2}$ approaches infinity, satisfying the decoding threshold $\gamma_{2}$. Meanwhile, the SINR for decoding $s_{1}$ converges to $ \frac{\alpha_{1}}{\alpha_{2}}$. To enhance the probability of successful decoding,  the ratio $ \frac{\alpha_{1}}{\alpha_{2}}$ is required to be greater than $\gamma_{1}$. Compared to conventional  fixed antenna-aided NOMA systems, PA-aided NOMA systems leverage LoS links and their inherent spatial flexibility to establish high-quality channels, enabling them to approach the required SINR at low power levels. When the SINR is large, the  security of the system is compromised, i.e.,  $\text{P}_{\text{sop}} = 1$. \label{cor 1} 
		%\end{cor}
		\begin{rem}\label{rem of cor}
			To ensure that the signal $s_{1}$ can be successfully decoded, the condition $\frac{\alpha_{1}}{\gamma_{1}} - \alpha_{2} > 0$ is required to be satisfied, i.e., $\frac{\alpha_{1}}{\alpha_{2}} >\gamma_{1} $. The increase of the transmit SNR decreases the security performance of PA-aided NOMA systems. 
		\end{rem}
		\section{Flexible Power Strategy}\label{section optimal problem}
		%\subsection{Flexible Power Mode aided Secure Transmission Strategy}
		The equal power model and the  proportional power model are the two conventional power model of PA systems.
		In the equal power model, the normalized power coefficients of two PAs are equal and less than or equal to $0.5$. This can be easily achieved by designing the PAs with varying coupling lengths using \eqref{power coefficient}. Although this model fully utilizes each antenna, it reduces the design flexibility in terms of antenna radiation power for NOMA system design, retaining only the same  superimposed signal power design at the BS as in conventional NOMA. In the proportional power model,  each PA is designed with the same coupling length. Consequently, the power emitted by each PA diminishes gradually along the waveguide, with the power radiated by subsequent PAs maintaining a fixed proportional relationship to that of the preceding ones. This characteristic reveals proportional power model cannot satisfy multiple scenarios. 
		
		Based on two conventional power models of PA, i.e., the equal power model and the proportional power model, a  flexible power strategy is proposed to address the confidential information leakage of  the PA-aided NOMA system with internal eavesdropping. Flexible power  strategy	achieves a precise power regulation, specifically realizing secure and reliable transmission from two dimensions of signal design and antenna radiation to resist potential internal eavesdropping. 
		In the flexible power strategy, $p_{1}$ and $p_{2}$ are strongly correlated, with $\epsilon_{2} = \left(1-F\sin^{2}\left(\kappa L_{1}\right)\right)\cdot F\sin^{2}\left(\kappa L_{2}\right)$. In a multi-PA system with a single waveguide, the PA closest to the BS has the highest priority in terms of power allocation. To  achieve secure communication, we formulate an optimal coupling length problem to satisfy the minimum SOP, which can be expressed as  
		
		\begin{subequations}
			\begin{align}
				\mathcal{P}_1: \quad &  \underset{L_{1},L_{2}}{\min}  \quad \text{P}_{\text{sop}}\left(L_{1}, L_{2}\right) \label{np11} \\
				&\text{s.t.} \quad \frac{\alpha_{1}}{\alpha_{2}}  > \gamma_{1}, \label{np12}\\
				& \quad \quad \min(\omega_{3},\omega_{4})>0,    \label{np13}\\
				&\quad \quad  \eta \left(\frac{\alpha_{1}}{\gamma_{1}}-\alpha_{2}\right)  >\eta\frac{\alpha_{2}}{\gamma_{2}}  , \label{np14}  \\
				&\quad \quad  0<L_{1},L_{2}\le \frac{\pi}{2\kappa}.\label{np15}
			\end{align}
		\end{subequations}
		Specifically, constraint \eqref{np12} ensures that the stronger signal $s_{1}$ can be successfully decoded by users from the superimposed signal. This satisfies the power allocation principle of NOMA. Constraints \eqref{np13} and \eqref{np14} guarantee that $\Omega_{1}$ and $\Omega_{2}$ in \eqref{y sop} are non-zero. Constraint \eqref{np15} defines the upper bounds of the coupling lengths $L_{1}$ and $L_{2}$. 
		%	\begin{table*}[!b]
			%	\caption{Feasible regions and optimal values of PAs' coupling lengths $L_{1}$ and $L_{2}$}
			%	\label{tab:pout}
			%	\centering
			%	\begin{tabular}{l||l||l}
				%		\hline\hline
				%		\textbf{Case} & \textbf{Feasible regions and optimal values of $L_{1}$} &\textbf{Feasible regions and optimal values of $L_{2}$}\\
				%		\hline\hline
				%		$\Omega_{1} = 1$, $\Omega_{2}$ = 1 &$ \Bigg( \frac{1}{\kappa}\arcsin\left(\sqrt{\frac{d^{2}+D^{2}/4}{BF\rho_{t}}}\right),$  & $\left( \frac{1}{\kappa}\arcsin\left(\sqrt{\frac{d^{2}+D^{2}/4}{\min\left(A,B\right)F\rho_{t}(1-Fs)} }\right),\frac{\pi}{2\kappa}
				%		\right)$ \\
				%		& $  \frac{1}{\kappa}\arcsin\left(\sqrt{ \min\left(\frac{1}{F}\left(1-\frac{d^{2}+D^{2}/4}{\min(A,B)F\rho_{t}}\right),1,\frac{d^2}{AF\rho_{t}}    \right)}\right)
				%		\Bigg)$ &\\
				%		\hline
				%		$\Omega_{1}$ = 1, $\Omega_{2}<1$  &  $\frac{1}{\kappa}\arcsin\left(\sqrt{\frac{d^{2}+D^{2}/4}{BF\rho_{t}}}\right)$       &$\frac{\pi}{2\kappa}$\\
				%		\hline
				%		$\Omega_{1} <1$, $\Omega_{2} = 1$ & $ \frac{1}{\kappa}\arcsin\left(\sqrt{\frac{d^2+D^{2}/4}{BF\rho_{t}}}\right)$ & $\left( \frac{1}{\kappa}\arcsin\left(\sqrt{	\frac{d^{2}+D^2/4}{\min\left(A,B\right)F\rho_{t}(1-F)\frac{d^{2}+D^{2}/4}{BF\rho_{t}}}}\right)   , \frac{\pi}{2\kappa}\right)$ \\
				%		\hline
				%		$\Omega_{1}<1$, $\Omega_{2}<1$	& $\frac{1}{\kappa}\arcsin\left(\sqrt{
					%			\frac{BF\rho_{t}-d^{2}}{2BF^{2}\rho_t}}\right)$ & $\frac{\pi}{2\kappa}$\\
				%		\hline\hline
				%	\end{tabular}
			%\end{table*}
			\begin{thm}\label{OPTHM}
				The optimal solution of problem $\mathcal{P}_{1}$ can be divided into four cases according to the maximum values of $\Omega_{1}$ and $\Omega_{2}$. %In case 1, both $\Omega_{1}$ and $\Omega_{2}$ can reach their maximum of $1$. In case 2, only $\Omega_{1}$ can reach its maximum of $1$. In case 3, only $\Omega_{2}$ can reach its maximum of  $1$. In case 4, both $\Omega_{1}$ and $\Omega_{2}$ have maximum values less than $1$. 
				Let $L_{i}^{m*}$ ($m\in\{1,2,3,4\}$) denote the optimal values of $L_{1}$ and $L_{2}$ along with their corresponding regions for each case $m$.  These are detailed as follows.
				\paragraph{Case 1} When $\Omega_{1}$ and $\Omega_{2}$ can reach their maximum of $1$, the feasible regions of $L_{1}$ and $L_{2}$ are given at the top of the next page, where $A = \frac{\eta\alpha_{2}}{\gamma_{2}}$, $B = \eta\left(\frac{\alpha_{1}}{\gamma_{1}}-\alpha_{2}\right)$, and $r =\sin^{2}(\kappa L_{1})$.
				\begin{figure*}
					\begin{align}
						L_{1}^{1*} &\in \Bigg( \frac{1}{\kappa}\arcsin\left(\sqrt{\frac{d^{2}+D^{2}/4}{BF\rho_{t}}}\right),   \frac{1}{\kappa}\arcsin\left(\sqrt{ \min\left(\frac{1}{F}\left(1-\frac{d^{2}+D^{2}/4}{\min(A,B)F\rho_{t}}\right),1,\frac{d^2}{AF\rho_{t}}    \right)}\right)
						\Bigg), \\
						L_{2}^{1*} &\in \left( \frac{1}{\kappa}\arcsin\left(\sqrt{\frac{d^{2}+D^{2}/4}{\min\left(A,B\right)F\rho_{t}(1-Fr)} }\right),\frac{\pi}{2\kappa}
						\right).
					\end{align}
					\hrulefill
				\end{figure*}
				\paragraph{Case 2} When $\Omega_{1}$ reaches its maximum of $1$ and the maximum of $\Omega_{2}$ is less than $1$, the optimal values of $L_{1}$ and $L_{2}$ are given by
				\begin{align}
					L_{1}^{2*}& = \frac{1}{\kappa}\arcsin\left(\sqrt{\frac{d^{2}+D^{2}/4}{BF\rho_{t}}}\right),\\
					L_{2}^{2*} &= \frac{\pi}{2\kappa}.
				\end{align}
				\paragraph{Case 3} When $\Omega_{2}$ reaches its maximum of $1$ and the maximum of $\Omega_{1}$ is less than $1$, the optimal value of $L_{1}$ and the feasible region of  $L_{2}$ are given by
				\begin{align}
					L_{1}^{3*}& = \frac{1}{\kappa}\arcsin\left(\sqrt{\frac{d^{2}+D^{2}/4}{BF\rho_{t}}}\right),\label{key}\\
					L_{2}^{3*} &\in \left( \frac{1}{\kappa}\arcsin\left(\sqrt{	\frac{d^{2}+D^2/4}{\min\left(A,B\right)F\rho_{t}\left( 1-\frac{F(d^{2}+D^{2}/4)} {BF\rho_{t}}\right)}}\right)   , \frac{\pi}{2\kappa}\right).
				\end{align}
				\paragraph{Case 4} When the maximum values of $\Omega_{1}$ and  $\Omega_{1}$  are less than $1$, the optimal values of $L_{1}$ and $L_{2}$ are given by
				\begin{align}
					L_{1}^{4*} &= \frac{1}{\kappa}\arcsin\left(\sqrt{
						\frac{BF\rho_{t}-d^{2}}{2BF^{2}\rho_t}}\right),\\
					L_{2}^{4*} &= \frac{\pi}{2\kappa}.
				\end{align}
				\begin{IEEEproof}
					See Appendix \ref{ProfT2}.
				\end{IEEEproof}
				%明天再说
				%The feasible regions and optimal values of $L_{1}$ and $L_{2}$ in four cases are shown at the top of the next page,  where \(A = \frac{\eta\alpha_{2}}{\gamma_{2}}\), and \(B=\eta\left(\frac{\alpha_{1}}{\gamma_{1}} - \alpha_{2}\right)\).	 
				\begin{rem}
					The optimal values of \(L_1\) in Cases 2 and 3 coincide with the lower bound of \(L_1^{1*}\) derived for Case 1. Meanwhile, the optimal values of \(L_2\) in Cases 2 and 4 are equivalent to the upper bound of \(L_2^{2*}\), which is \(\frac{\pi}{2\kappa}\). In Case 1, feasible regions of  $L_{1}$ and $L_{2}$ are derived. The users can adjust the coupling lengths of PAs with this region to achieve zero outage. In Cases 2, 3, and   4, the optimal values of coupling lengths are derived, respectively. Since the minimum value of SOP may be greater than zero, the secrecy outage event may occur with a non-zero probability. %However, using the optimal lengths in Table \ref{tab:pout}, the users can still achieve  a reliable and secure communication under these suboptimal conditions.
				\end{rem} 
			\end{thm}
			\begin{rem}\label{rem4}
				The equal power model and the proportional power model cannot address  the issue of  PA-aided NOMA systems. In the equal power model, both PAs radiate the same power.
				As the channel conditions for both users are identical, the probability of the public information user decoding confidential information is the same as that of the confidential information user. The equal power model exposes the NOMA systems to a high risk of confidential information leakage. In  the proportional power model,  PA-2 radiates less power, which may not satisfy $U_{2}$'s uninterrupted transmission but decreases the security  performance of system. To address these issues, the flexible power strategy can allocate the radiation power by adjusting  coupling lengths of PA, in which PA-1 radiates less power to public information user $U_{1}$ and PA-2 radiates more power to confidential information user $U_{2}$. The proposed flexible power strategy ensures both $U_{1}$ and $U_{2}$ can decode their own information and simultaneously prevents $U_{1}$ from intercepting the confidential information of $U_{2}$.
			\end{rem}
			\begin{rem}\label{rem5}
				PA-1 serves the public information user since they are closer to the BS. The remaining waveguide power is radiated to $U_{2}$ by PA-2, which is $\rho_{t}(1-F\sin^2(\kappa L_{1}))F\sin^2(\kappa L_{2})$. By radiating all of the waveguide's power to $U_{1}$, PA-1 can stop $U_{2}$ from transmitting. Therefore, PA-1 to service the user of confidential information is a viable deployment for both users and PAs.
			\end{rem}
			
			\section{Simulation Results}\label{section 4}
			In this section, we conduct a comprehensive evaluation of the security performance of the PA-aided NOMA system. The simulation results show the accuracy of the theoretical analysis and illustrate the impacts of key parameters. Unless otherwise stated, we set the carrier frequency $f_{c} = 28$ GHz, the height of waveguide $d = 3$ m, the speed of light $c = 3\times10^{8}$ m/s,  the effective refractive index $n_{eff} = 1.4$ \cite{Pozar2011}, and coupling coefficient $\kappa = 100$ m$^{-1}$. The coupling  efficiency is fixed at $F = 1$\footnote{From the above formulas, it can be inferred that the power emitted by a PA can be regulated by adjusting the coupling length $L$. When the waveguide and the separate dielectric possess  the same effective refractive index, the coupling efficiency $F$ attains  its maximum value of 1, enabling the radiation of full power from a single PA with a coupling length of $\pi/(2\kappa)$.},  with decoding thresholds $\gamma_{1} = 10$ dB and $\gamma_{2} =15$ dB.  Additionally, we set the power allocation coefficients  $\alpha_{1} = 0.01$ and $\alpha_{2} = 0.99$, $D_{1} = D_{2} = 10$ m, and the dimensions of the areas as $D\times D = 10\times 10 \medspace \text{m}^2$. The simulated performance is obtained by performing Monte Carlo simulations over \({10}^{6}\) realizations.  
			\begin{figure}[t]
				\centering
				\includegraphics[width = 3.8 in]{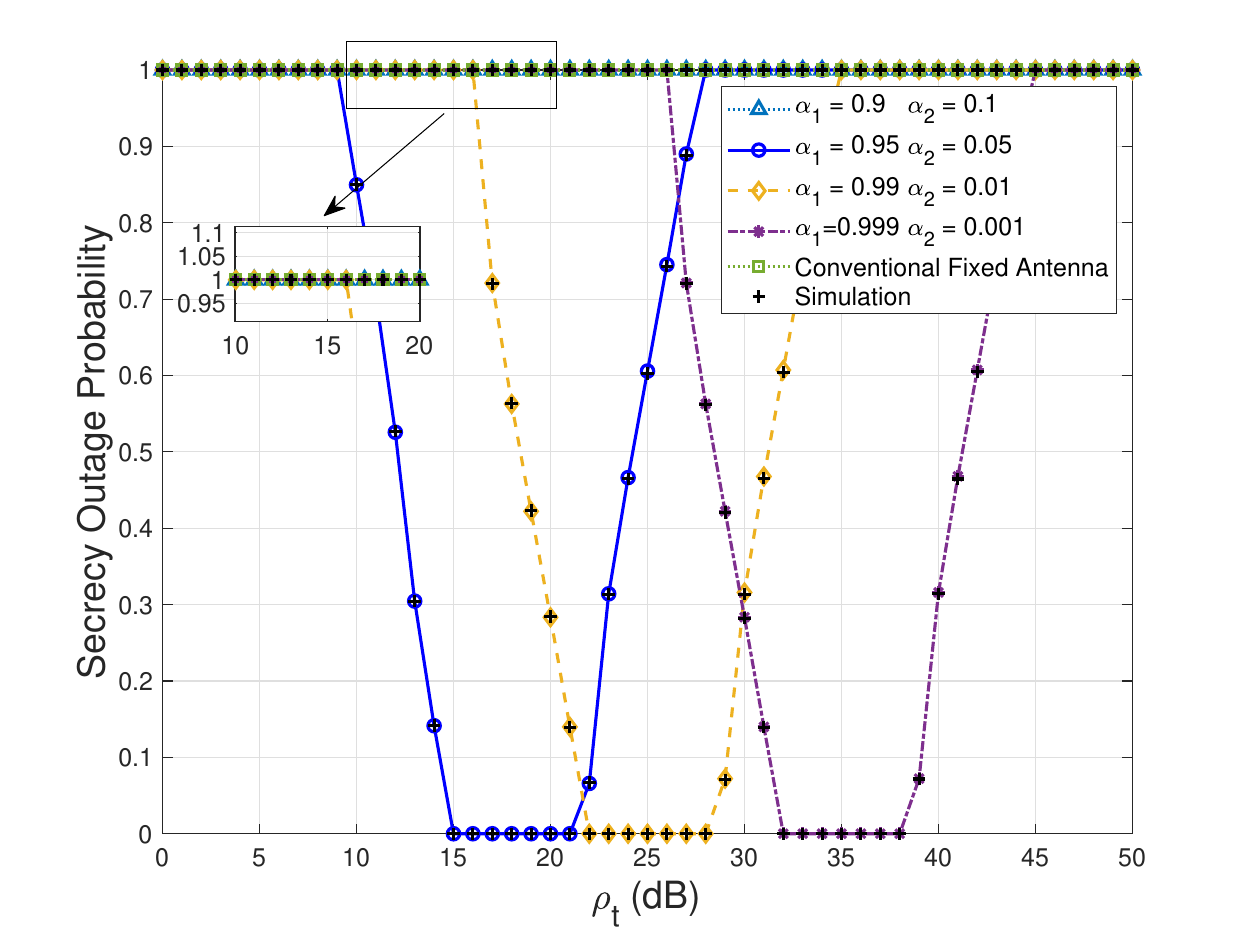}
				\caption{Secrecy outage probability versus $\rho_{t}$  for different pairs of signal power allocation coefficients with $L_{1} =1\times 10^{-3}$ and $L_{2} = 1.57\times 10^{-2}$.}
				\label{sim:1}
			\end{figure}
			\begin{figure}[t]
				\centering
				\subfigure[$\alpha_{1} = 0.9$ and $\alpha_{2} = 0.1$]{
					%\label{fig:side_c1}
					\includegraphics[scale=0.19]{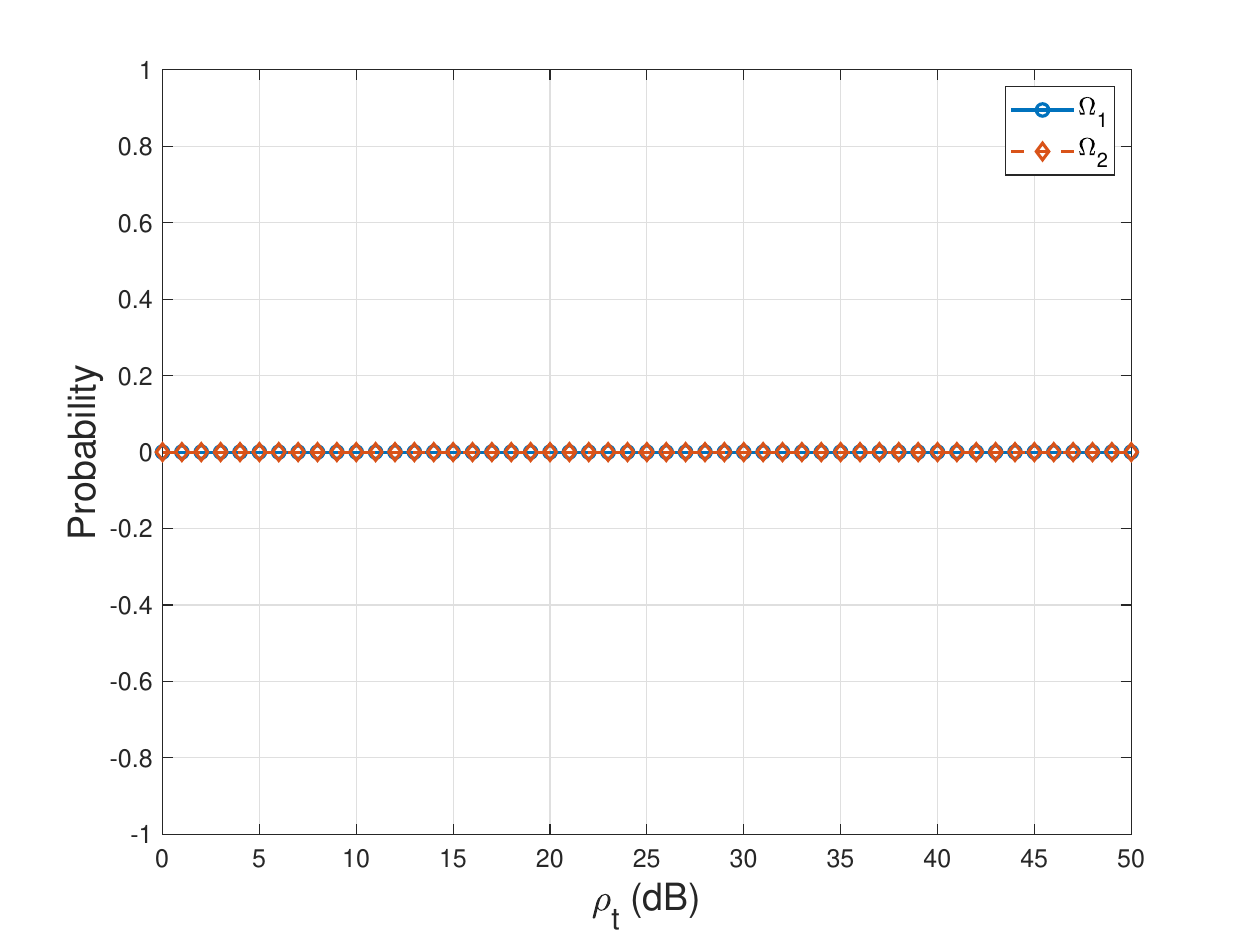}
				}
				\subfigure[ $\alpha_{1} = 0.95$ and $\alpha_{2} = 0.05$]{
					%\label{fig:side_c2}
					\includegraphics[scale=0.19]{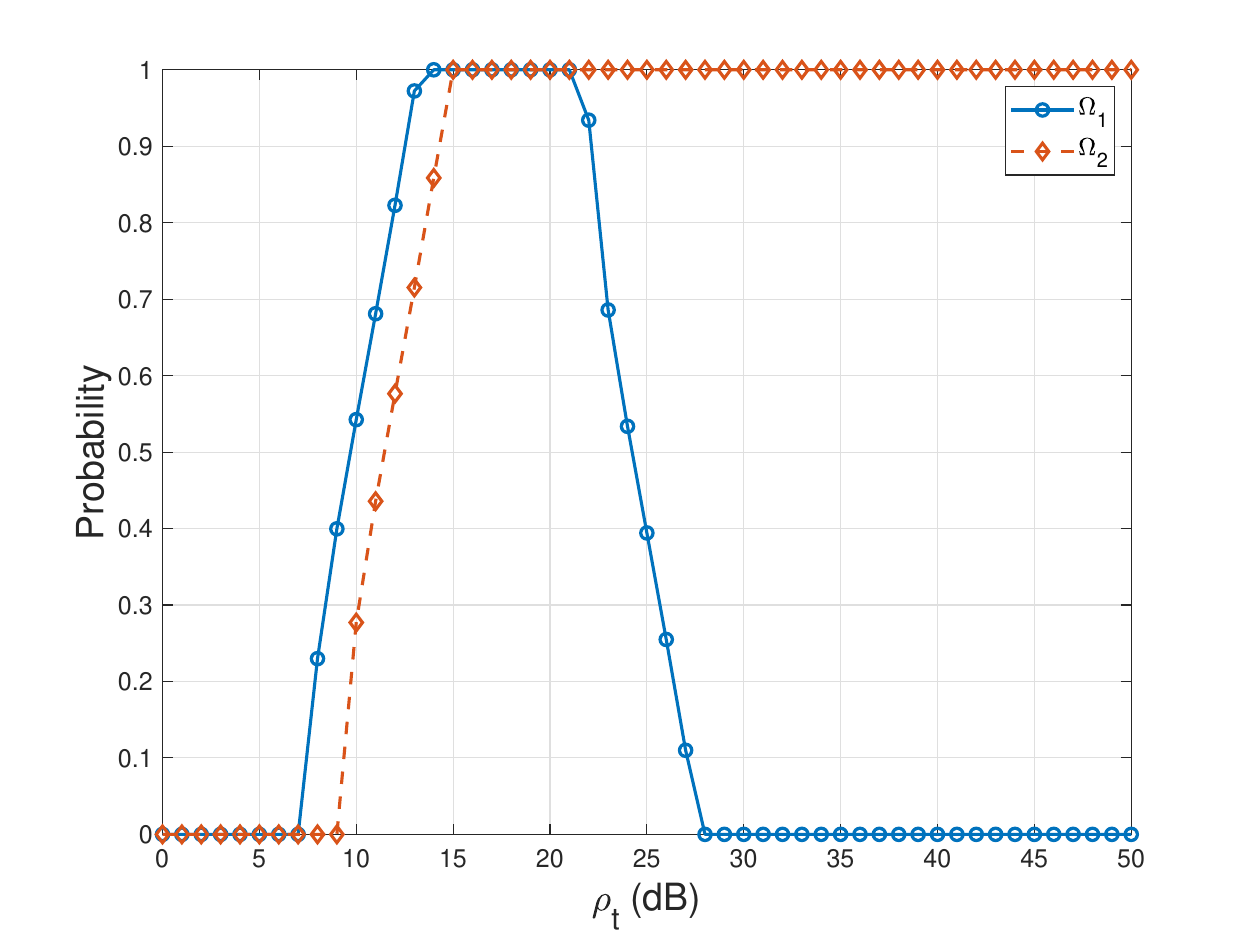}
				}
				\subfigure[$\alpha_{1} = 0.99$ and $\alpha_{2} = 0.01$]{
					%\label{fig:side_c1}
					\includegraphics[scale=0.19]{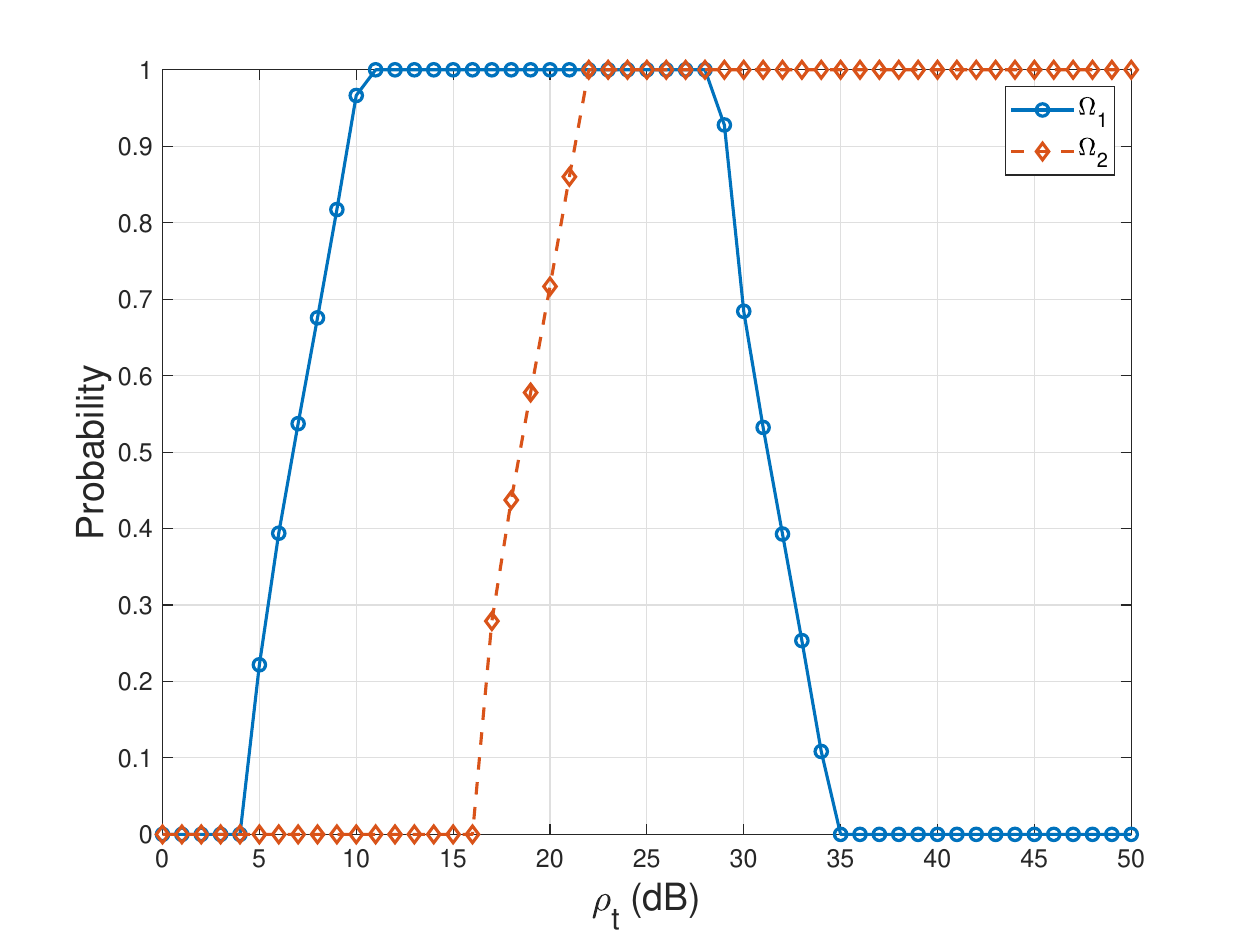}
				}
				\subfigure[$\alpha_{1} = 0.999$ and $\alpha_{2} = 0.001$]{
					%\label{fig:side_c1}
					\includegraphics[scale=0.19]{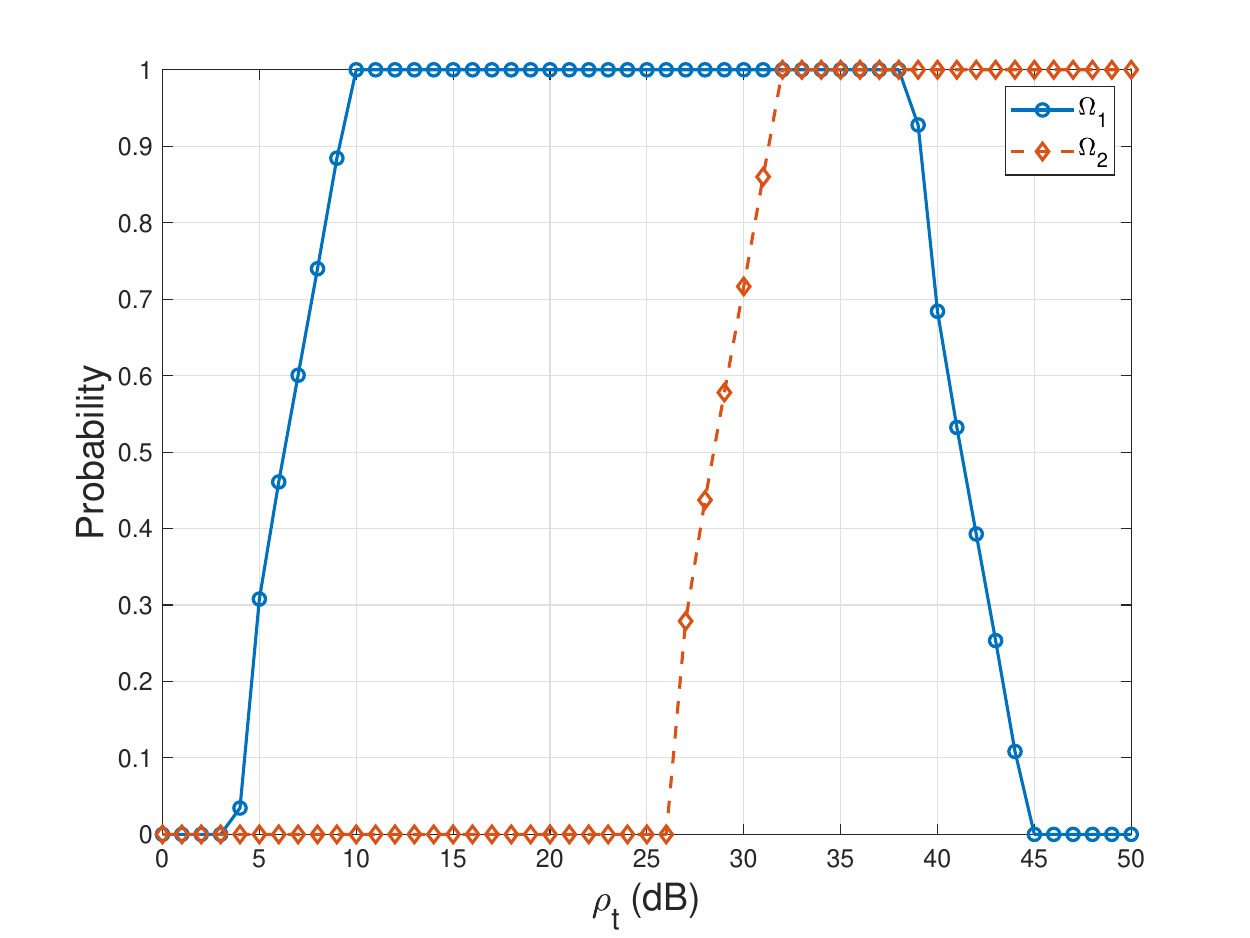}
				}
				\caption{The probability that $U_{1}$ achieves reliable transmission  and the information of $U_{2}$ is not leaked ($\Omega_{1}$), and the probability that $U_{2}$ achieves reliable transmission ($\Omega_{2}$) versus transmit SNR $\rho_{t}$ for different pairs of signal power allocation coefficients with $L_{1} =1\times 10^{-3}$ and $L_{2} = 1.57\times 10^{-2}$.}
				\label{2.1}
			\end{figure}
			
			Fig \ref{sim:1} illustrates the SOP versus transmit SNR $\rho_{t}$ for different signal power allocation coefficients. It can be observed that the numerical results validate the correctness of the theoretical analysis. The SOP of a conventional fixed antenna-aided NOMA system is considered as a benchmark which shows the superiority of PA-aided NOMA systems in enhancing security. Moreover, as $\rho_{t}$ increases,  the SOP firstly decreases  maintains at zero over a certain interval, then increases to 1.
			However, when $\alpha_{1} = 0.9$ and $\alpha_{2} = 0.1$, the secrecy outage occurs with the probability of 1. The reason is that the ratio of $\alpha_{1}$ to $\alpha_{2}$ being lower than $\gamma_{1}$  confirms the conclusion in Remark \ref{rem of cor}. The decreased $\alpha_{2}$ does not improve the security performance of PA-aided NOMA systems, rather, it achieves $\text{P}_{\text{sop}}=0$ in the higher SNR range. Because, reducing the value of $\alpha_{2}$ decreases the risk of confidential information leakage while increasing the transmission outage probability of $U_{2}$. Additionally, the asymptotic results match the simulation results at high SNR, confirming the accuracy of our asymptotic analysis.
			
			Fig. \ref{2.1} illustrates the probability  ($\Omega_{1}$) that $U_{1}$ achieves a  reliable transmission and the information of $U_{2}$ is not leaked and the probability ($\Omega_{2}$) that $U_{2}$ achieves a reliable transmission. The increases of $\Omega_{1}$ and $\Omega_{2}$ mean the improvement of reliability and security performance of PA-aided NOMA systems. For the curves corresponding to different $\alpha_{1}$ and $\alpha_{2}$, the overall trends are similar.
			Specifically, as $\rho_{t}$ increases, $\Omega_{1}$ first rises, maintains at its maximum value of $1$ over a certain interval, then decreases to zero. Reducing the value of $\alpha_{2}$ expands the range of  SNR values over which $\Omega_{1} = 1$. This is because security performance is the domain factor with high SNR; the decrease in $\alpha_{2}$ increases the difficulty for the public information user to successfully decode $s_{2}$. However, the confidential information user requires a high SNR to decode $s_{2}$ with a small power allocation coefficient for $s_{2}$. 
			The SNR required for $\Omega_{2}$ to reach its maximum value is higher than that for $\Omega_{1}$. As SNR increases, $\Omega_{2}$ begins to rise during the decline phase of $\Omega_{1}$, and the peak portions of $\Omega_{1}$ and $\Omega_{2}$ do not overlap. This explains why the increase of SNR decreases the SOP achievable by the PA-NOMA system.
			Simulation results validate the conclusions drawn in Remark \ref{thm3}.
			
			%\begin{figure}[t]
			%	\centering
			%	\includegraphics[width = 3.8 in]{sim_pic/pdf/alpha_change}
			%	\caption{The probability that $U_{1}$ achieves reliable transmission  and the information of $U_{2}$ is not leaked ($\Omega_{1}$), and the probability that $U_{2}$ achieves reliable transmission ($\Omega_{2}$) versus transmit SNR $\rho_{t}$ for different pairs of signal power allocation coefficients with $L_{1} =1\times 10^{-3}$ and $l_{2} = 1.57\times 10^{-2}$.}
			%	\label{2.1}
			%\end{figure}

			\begin{figure}[t]
				\centering
				\includegraphics[width = 3.8 in]{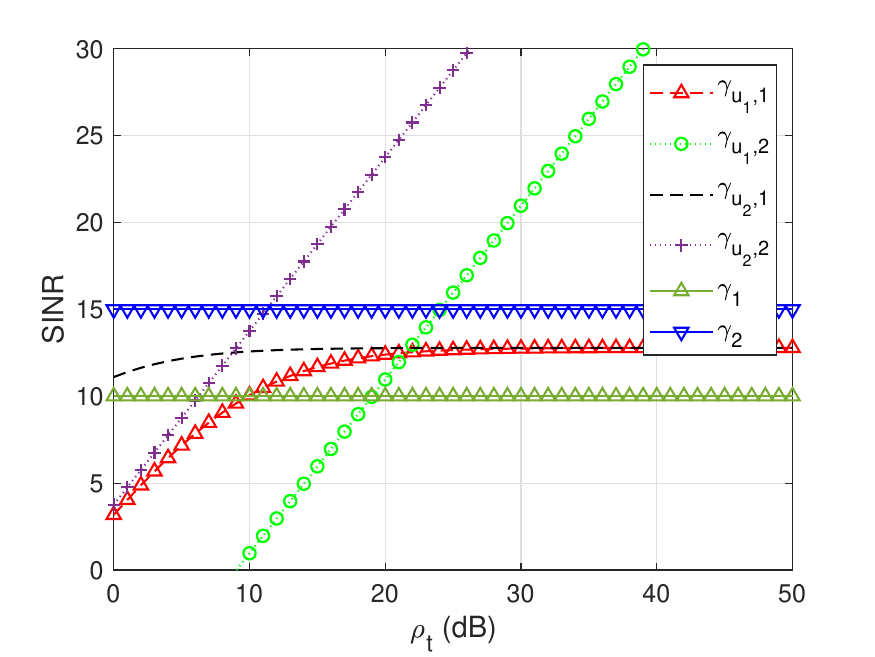}
				\caption{Signal decoding SINRs of users versus $\rho_{t}$ with $\alpha_{1} = 0.99$, $\alpha_{2} = 0.01$, $L_{1} =1\times 10^{-3}$, and $L_{2} = 1.57\times 10^{-2}$.}
				\label{sim:0}
			\end{figure}
			
			\begin{figure}[t]
				\centering
				\includegraphics[width=3.8in]{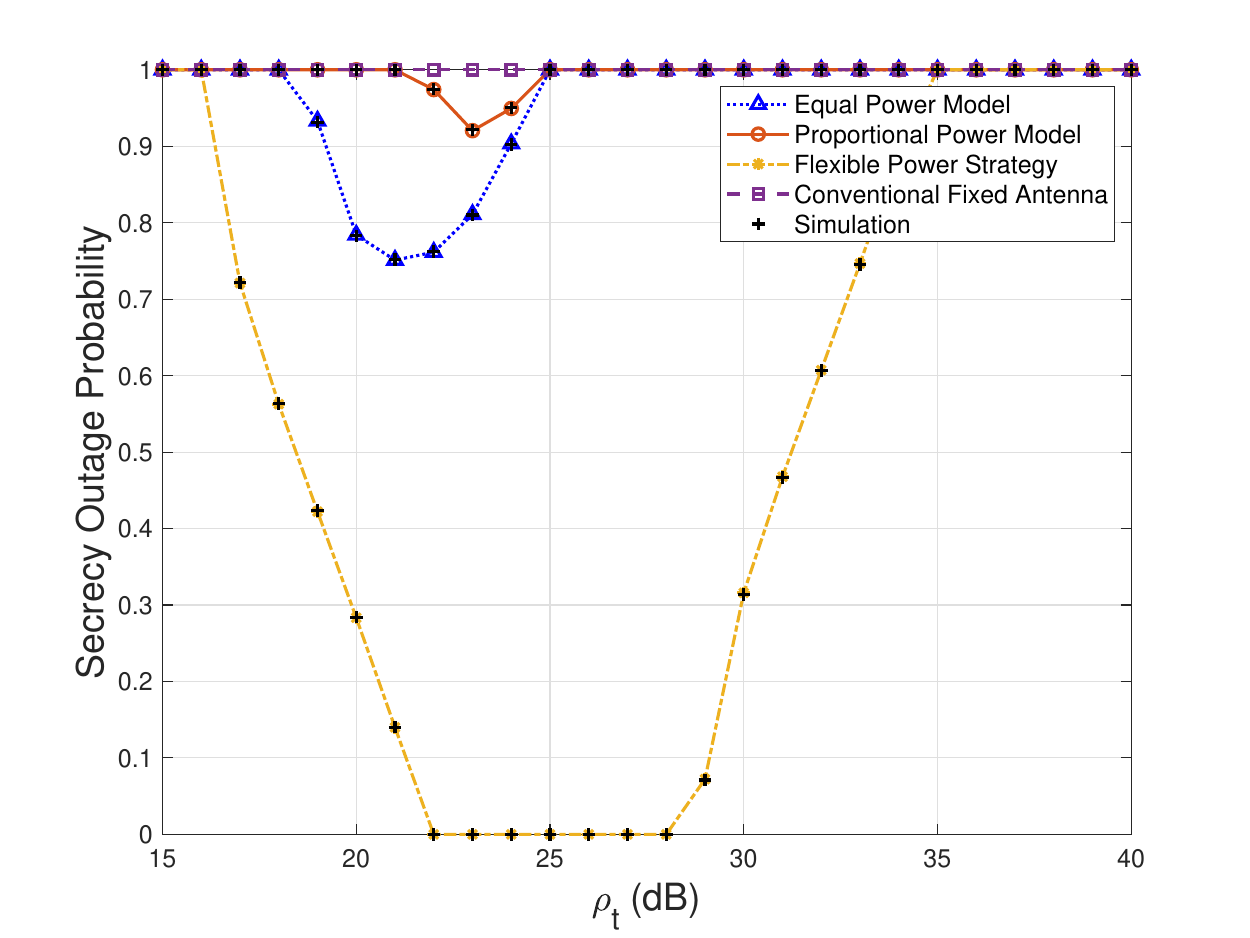}
				\caption{
					Comparison among the secrecy outage probability in the flexible power strategy, the secrecy outage probabilities in the equal power model and the proportional power model, and the secrecy outage probability in the conventional fixed antenna aided NOMA system.}
				%Secrecy outage probability of versus $\rho_{t}$ for various power models.}
			\label{compare of mode}
		\end{figure}
		
		\begin{figure}
			\centering
			\subfigure[The impact of changing the side length of $U_{1}$'s activity cell on the SOP with the side length of $C_{2} = 10$.]{
				\label{fig:side_c1}
				\includegraphics[width=3.8in]{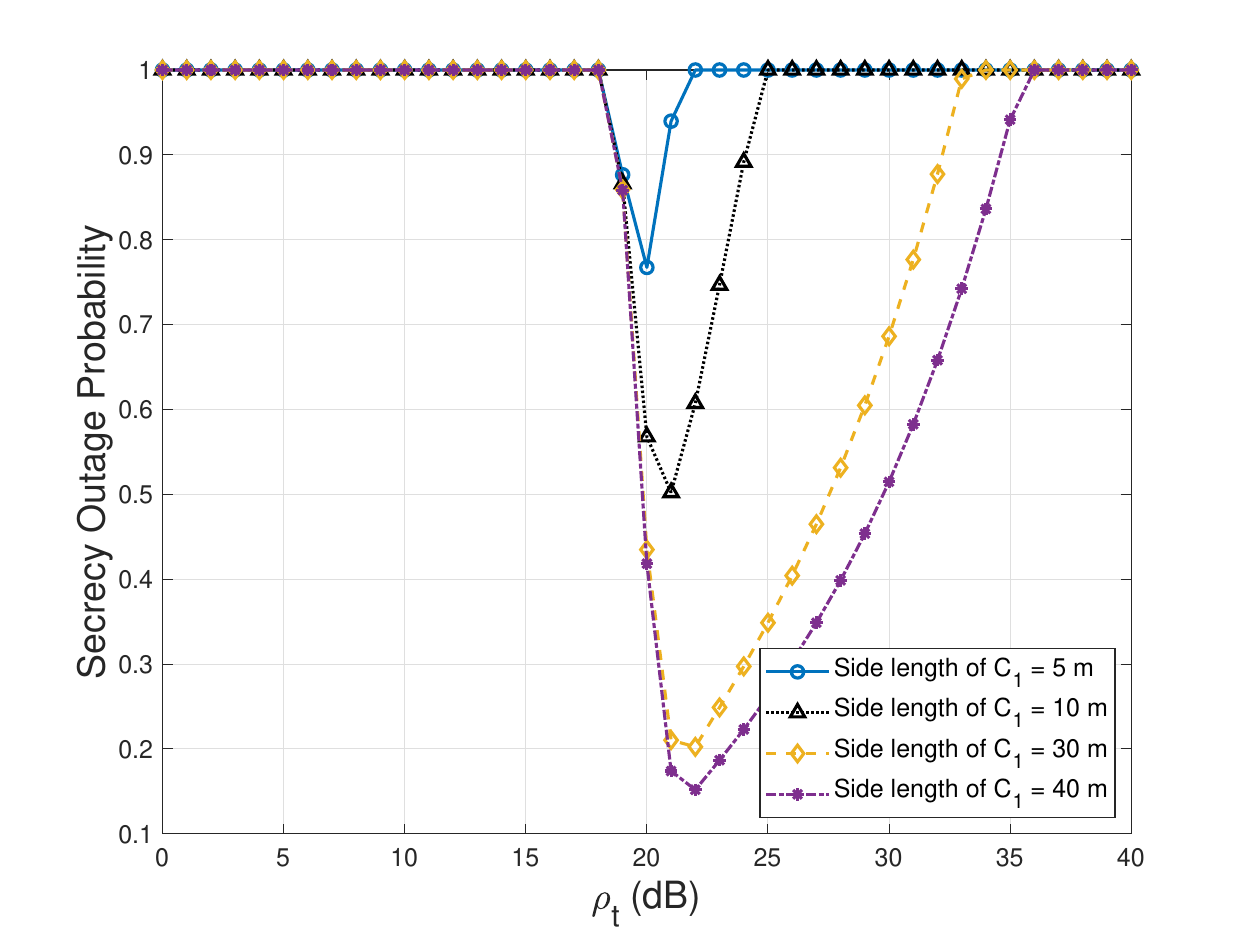}}
			\subfigure[The impact of changing the side length of $U_{2}$'s activity cell on the SOP with the side length of $C_{1} = 10$.]{
				\label{fig:side_c2}
				\includegraphics[width=3.8in]{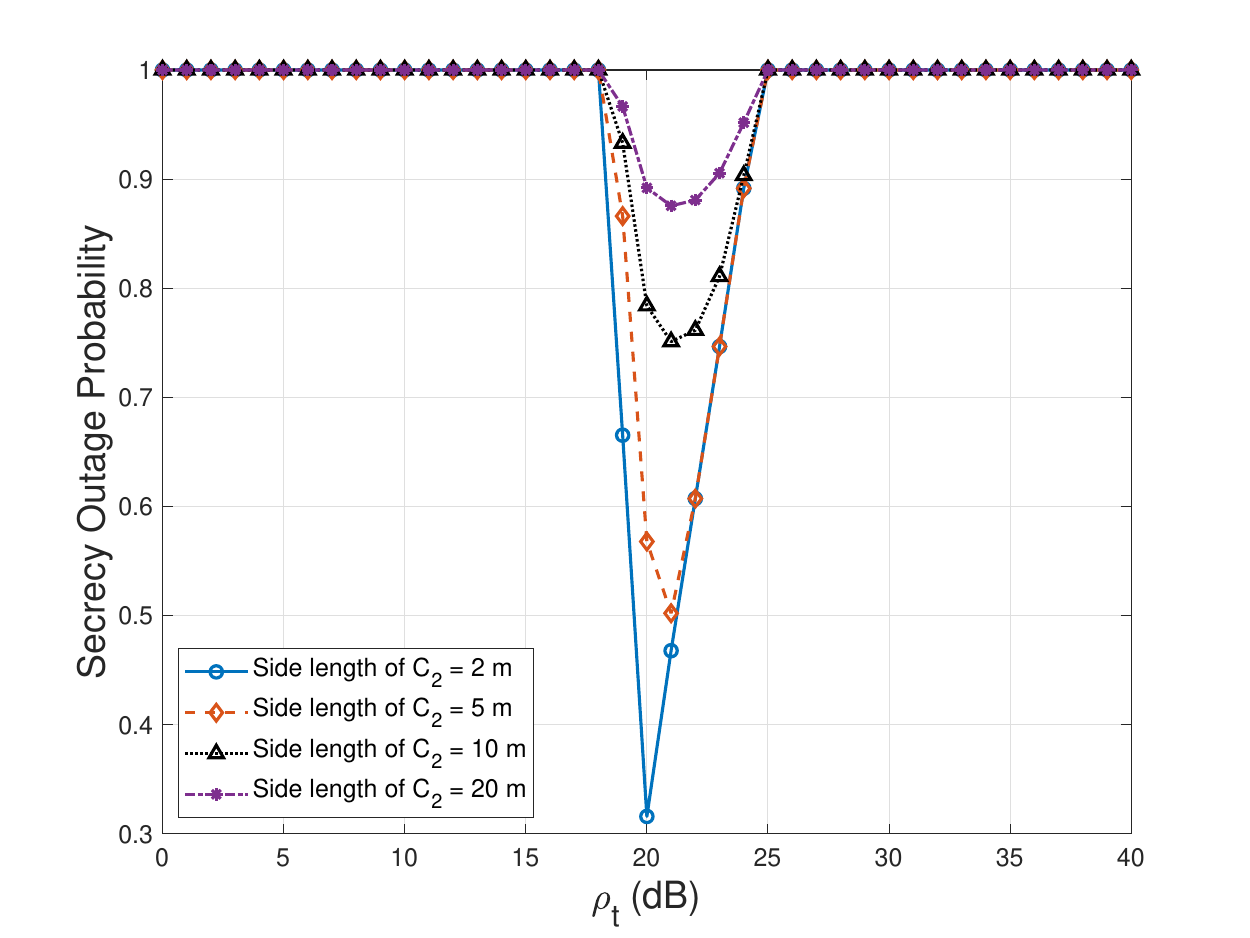}}
			\caption{Secrecy outage probability versus $\rho_{t}$ for different side lengths of cells with $\epsilon_{1} = \epsilon_{2} = 0.5$.}\label{side_CELLS}
		\end{figure}

		\begin{figure}[t]
			\centering
			\includegraphics[width=3.8in]{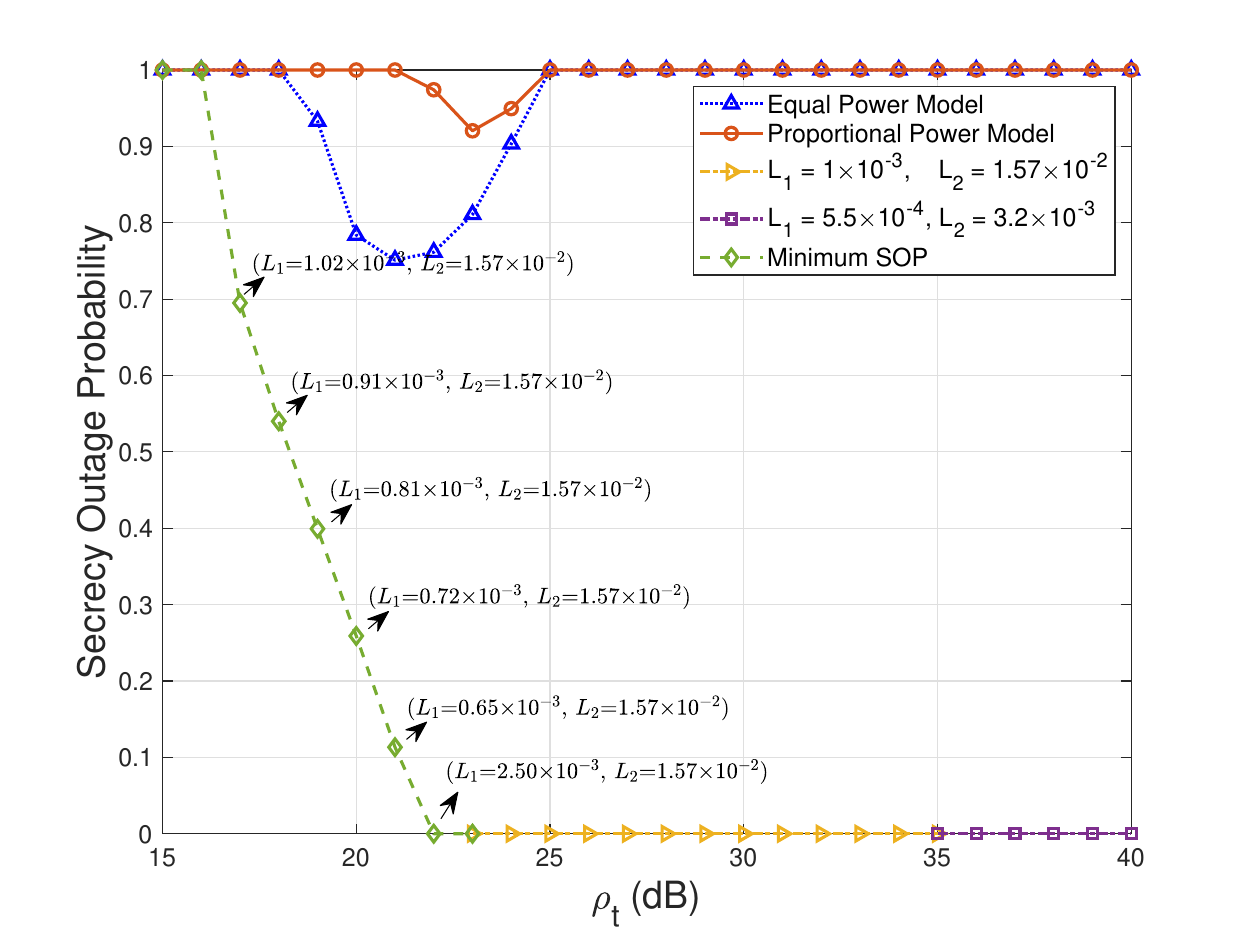}
			\caption
			{Secrecy outage probability in the
				flexible power  strategy versus $\rho_{t}$.}
			\label{sim:7}
		\end{figure}
		\begin{figure}[t]
			\centering
			\includegraphics[width=3.8in]{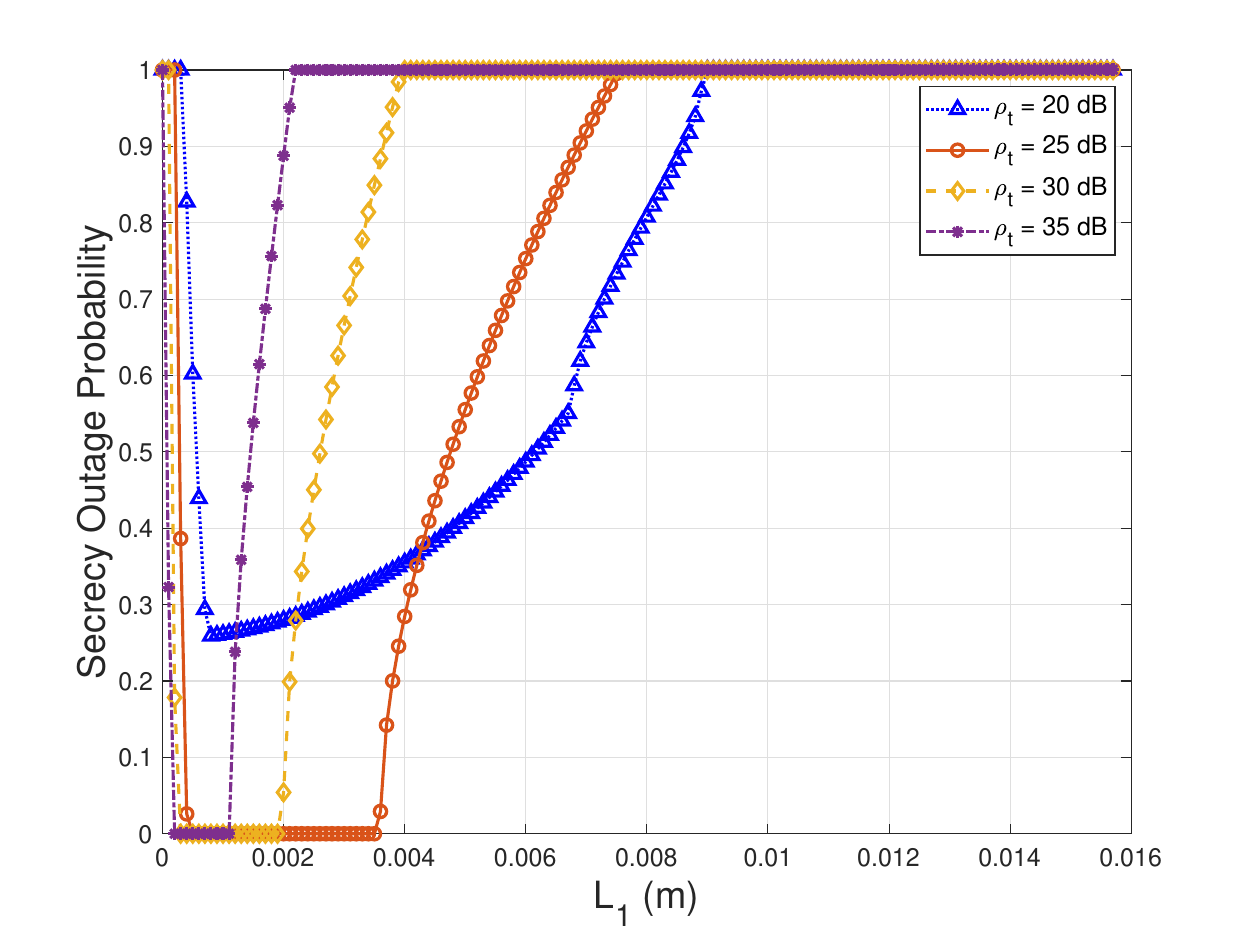}
			\caption
			{Secrecy outage probability versus coupling length of PA-1.}
			\label{sim:8}
		\end{figure}
		Fig. \ref{sim:0} shows the received SINRs at users. To guarantee the effective decoding of $s_{1}$, the  power coefficient ratio $\frac{\alpha_{1}}{\alpha_{2}}$ is greater than $\gamma_{1}$. In the equal power model, users decode the same signals with the same SINRs, i,.e., $\gamma_{u_{1},1} = \gamma_{u_{2}}$ and $\gamma_{u_{2},1} = \gamma_{u_{2},2}$. In the proportional power model, the security performance is worse than that in the equal power model, as  required for \(U_2\) to decode the confidential information is lower than that for \(U_1\).
		Conventional power models for PA are limited in their ability to improve the security of PA-aided NOMA systems as  they cannot adjust the  power flexibly.  In the flexible power strategy, the PA-aided NOMA systems can adjust the radiated power of PAs increase the SINR required for $U_{2}$ to decode the confidential information at low transmit SNR, while decreasing the SINR required for \(U_1\) to decode the confidential information, i.e., \(\gamma_{u_1,2} \ll \gamma_{u_2,2}\).
		
		In Fig. \ref{compare of mode}, the SOP of the PA-aided NOMA system in the flexible power strategy is contrasted with the SOPs of the PA-aided NOMA systems in the two conventional power models and the conventional fixed antenna-aided NOMA system. The results validate the findings in Remarks \ref{rem3} and \ref{rem4}, demonstrating that in the flexible power strategy, PAs fulfill the transmission requirements of both authorized users while reducing the risk of confidential information leakage. 
		When $\rho_t = 21$ dB, the system's SOP in the equal power model reaches a minimum of $0.75$ at $L_{1} = \frac{\pi}{4\kappa}$ and $L_{2} = \frac{\pi}{2\kappa}$. For the proportional power model, we set $L_{1} = L_{2} = \frac{\pi}{4\kappa}$,  where $\epsilon_{2}$ attains its maximum value of $0.25$ and the system achieves a minimum SOP of $0.92$ when $\rho_{t} = 23$. Notably, the conventional power models allocate  more (or equal) power to $U_{1}$,  which increases the risk of information leaks. In contrast, the flexible power strategy at \(L_{1} = \frac{\pi}{14\kappa}\) and \(L_{2} = \frac{\pi}{2\kappa}\) allocates more power to \(U_{2}\) and less to \(U_{1}\), resulting in the SOP remaining at 0 for transmit powers between $22$ dB and $28$ dB.
		
		Fig. \ref{fig:side_c1} illustrates the effect of the side length of $C_{1}$ on the SOP of system. Using the curve for \(C_1\)’s side length of 10 as a benchmark, it can be observed that reduction in the public information user’s activity cell (\(C_1\)) leads to an increase in the SOP.
		Correspondingly, Fig. \ref{fig:side_c2} represents how the side length of \(C_2\) influences the SOP.  As $C_{2}$ expands, the SOP increases progressively. Specifically, an increase in activity cell leads to a greater distance between the user and the PA, resulting in a decrease in channel strength. Conversely, a smaller activity cell yields higher channel strength. When the activity cell of the public information user increases, the radiated power of the superimposed signal received by the user decreases significantly. Since \(\alpha_1 \gg \alpha_2\), $U_{1}$ can successfully decode \(s_1\), but the SINR required for $U_1$ to decode \(s_2\) is far below the threshold. Consequently, the SOP decreases, and the system security is enhanced. Correspondingly, when the activity cell \(C_2\) increases, the signal power received by $U_{2}$ attenuates drastically, preventing the decoding of \(s_2\). This leads to a decline in the communication reliability of $U_{2}$ and, in turn, an increase in the SOP.

		Fig. \ref{sim:7} shows the SOPs versus $\rho_{t}$ under the flexible power  strategy. With the help of Theorem \ref{OPTHM}, we design the optimal values of $L_{1}$ and $L_{2}$ within the SNR range between $15$ dB and $40$ dB to achieve the minimum SOP.  When the transmit SNR is below 15 dB, the SINR required for \(U_2\) to decode \(s_2\) cannot reach the decoding threshold, thus the SOP remains 0. The simulation results demonstrate that by adjusting the coupling lengths of PA, the flexible power model can satisfy system security requirements under various SNR conditions.  The results show that the conventional power models fail to adapt to the dynamic changes of \(\rho_{t}\).  The flexibility of  PA is represented in two aspects. On the one hand, dynamic position adjustment can minimize
		the transmission distance, thereby reducing path loss. On the
		other hand, precise control of amplification power can enhance
		system security. It can thus be concluded that the PA-aided
		NOMA system introduces new degrees of freedom for PLS
		through the design of normalized power coefficients and the
		optimized deployment of positions.
		%The proposed strategy effectively improve the security performance of PA-aided NOMA systems, especially in scenarios where public information users are close to the BS. PA can dynamically adjust its transmission power by flexibly modifying the coupling length to cope with harsh eavesdropping scenarios, thereby significantly expanding the non-outage interval of the system. Flexibility is a core characteristic of PAs, which is manifested in two aspects. On the one hand, dynamic position adjustment can minimize the transmission distance, thereby reducing path loss. On the other hand, precise control of amplification power can enhance system security. It can thus be concluded that the  PA-aided NOMA system introduces new degrees of freedom for PLS through the design of normalized power coefficients and the optimized deployment of positions.

		Fig. \ref{sim:8} shows the SOPs versus $L_{1}$ with  $L_{2} = 1.57 \times 10^{-2}$ m.
		Table \ref{tab:coupling length} shown at the top of the next page presents the coupling lengths through theoretical analysis and simulation when the minimum $\text{P}_{\text{sop}}$ is achieved with various SNRs. The simulation results verify the accuracy of the theoretical analysis, with errors attributed to limitations in calculation precision and the step size setting during the simulation process. When $17\le\rho_{t}\le21$ dB, the minimum value of $\text{P}_{\text{sop}}$ is greater than $0$ as shown in Table \ref{tab:coupling length}, the maximum value of \(\Omega_{1}\) is 1, and the maximum value of \(\Omega_{2}\) is less than 1.  For the other three SNR values, a suitable set of \(L_{1}\) and \(L_{2}\) can be found to achieve $\text{P}_{\text{sop}} = 0$. The numerical results demonstrate that the proposed optimization scheme can effectively guide the design of PA, thereby improving system security.

		\begin{table*}[!t]
			\caption{Theoretical vs. simulated coupling length}
			\label{tab:coupling length}
			\centering
			\begin{tabular}{l||l||l||l}
				\hline\hline
				\textbf{$\rho_{t}$ (dB)} & \textbf{Theoretical $L_{1}$ (m)} &\textbf{Simulated $L_{1}$ (m)} & min $\text{P}_{\text{sop}}$\\
				\hline\hline
				17 & 1.02$\times$ 10 $^{-3}$&1.1$\times$10 $^{-3}$&0.6950\\
				\hline
				18 & 9.1$\times$10$^{-34}$& 1$\times$10$^{-3}$ & 0.5401 \\
				\hline
				19 & 8.1$\times$10$^{-4}$& 9$\times$10$^{-4}$& 0.3993\\
				\hline
				20& 7.25$\times$10$^{-4}$ & 8$\times$10$^{-4}$&0.2590\\
				\hline
				21& 6.46$\times$10$^{-4}$ & 7$\times$10$^{-4}$&0.1133\\
				\hline
				22 & (5.76$\times$10$^{-4}$,2.61$\times$10$^{-3}$)& [6$\times$10$^{-4}$,2.5$\times$10$^{-3}$]& 0\\
				\hline
				23 & (5.13$\times$10$^{-4}$,4.58$\times$10$^{-3}$)& [6$\times$10$^{-4}$,4.5$\times$10$^{-3}$]& 0\\
				\hline\hline
			\end{tabular}
		\end{table*}
		\section{Conclusion}\label{section 5}
		In this paper, we investigated a PA-aided NOMA system with internal eavesdropping, where the near public information user ($U_1$) is an internal eavesdropper who eavesdrops on the far confidential information user ($U_2$). To evaluate the security of the system, we derived the  exact closed-form expressions for the SOPs and analyzed the asymptotic behaviors to gain valuable insights.  Additionally, to strengthen the security of NOMA systems with internal eavesdroppers, we proposed a  flexible power strategy to achieve secure transmission. This strategy features adaptive power allocation, by optimizing coupling lengths of the PA, it allocates more transmission power to $U_{2}$ and less to $U_{1}$. This design specifically addresses the issue that the conventional fixed antenna systems and PA systems (in the equal power model and the proportional power model)  have a high risk of confidential information leakage when  the internal eavesdropper is near the BS.
		The results demonstrate that setting small power allocation coefficients to confidential information while allocating more power to the confidential information user via coupling length adjustment is an effective strategy to improve the system’s security performance. When the range of the users' activity cells is smaller than the maximum reliable transmission distance, while the range of the eavesdropper's activity cell exceeds the maximum eavesdropping distance, the PA-aided NOMA system can effectively mitigate the risk of confidential information leakage.
		
		%Moreover, in order to direct the deployment of users and PAs, we define the maximum distance at which $U_{1}$ can successfully eavesdrop on and the maximum reliable transmission distance for users.

		\begin{appendices}  
			\section{Proof of Lemma 1}\label{ProfL1}
			To derive the closed-form expressions of $\Omega_{1}$ in $\eqref{varOmega1}$, we consider the relationship between feasible range of  $y_{1}^{2}$, i.e., $0\le y_{1}^{2}\le\frac{D^{2}}{4}$ and the magnitudes of $\omega_{1}$ and $\omega_{2}$. The proof of $\Omega_{1}$ is given by
			\begin{align}\label{pr O1}
				\Omega_{1} =& \text{Pr}\left[\omega_1<y_{1}^{2}\le \omega_{2}\right]\notag\\
				=&\frac{1}{D}\Bigg[\mathbb{I}\left(\omega_{1}\le0,\omega_{2}\ge\frac{D^{2}}{4}\right)\int_{-\frac{D}{2}}^{\frac{D}{2}}dy_{2}\notag\\
				&+\mathbb{I}\left(\omega_{1}\le0,0<\omega_{2}<\frac{D^{2}}{4}\right)\int_{-\sqrt{\omega_{2}}}^{\sqrt{\omega_{2}}}dy_{2}\notag\\
				&+\mathbb{I}\left(0<\omega_{1}<\omega_{2}<\frac{D^{2}}{4}\right)\left(\int_{-\sqrt{\omega_{2}}}^{-\sqrt{\omega_{1}}}dy_{1} + \int^{\sqrt{\omega_{2}}}_{\sqrt{\omega_{1}}}dy_{1}  \right)\notag\\
				&+\mathbb{I}\left(0<\omega_{1}<\frac{D^{2}}{4},\omega_{2}\ge\frac{D^{2}}{4}\right)\left(\int_{-\frac{D}{2}}^{-\sqrt{\omega_{1}}}dy_{1} + \int^{\frac{D}{2}}_{\sqrt{\omega_{1}}}dy_{1} \right)
			\Bigg].
			\end{align}
			By performing algebraic transformations on \eqref{pr O1}, we derive $\Omega_{1}$ as shown in \eqref{varOmega1}, thus completing the proof.
			\section{Proof of Theorem 2}\label{ProfT2}
			According to \eqref{y sop}, the optimization problem $\mathcal{P}_{1}$ in \eqref{np11} aimed at minimizing SOP is equivalent to maximizing  $\Omega_{1}\cdot\Omega_{2}$ the probability of the valid secure transmission event. The equivalent relation is given by 
			\begin{align}
				\min\text{P}_{\text{sop}} \Leftrightarrow \max\text{Pr}[\mathbf{\varepsilon}],
			\end{align}
			where $\varepsilon = \left\{    \omega_{1}<y_{1}^{2}\le\omega_{2}, y_{2}^{2}\le\min\left(\omega_{3},\omega_{4}\right)\right\}$ denotes the event that  $U_{1}$ decodes $s_{1}$ successfully but fails to decode $s_{2}$, and $U_2$ decodes both $s_{1}$ and $s_{2}$ successfully. Since the maximum value of $\text{Pr}[\varepsilon]$ is not guaranteed to be 1, we denote its maximum value as $M \in (0,1]$. Based on the former analysis of \eqref{varOmega1} \eqref{Omega2}, the feasible region  for $L_{1}$ and $L_{2}$ are divided into four cases.    Notably, let \(r = \sin^2(\kappa L_1)\) and \(t = \sin^2(\kappa L_2)\). Let \(r^{j*}\) and \(t^{j*}\) denote the corresponding optimal values of $r$ and $t$ in different cases, respectively.%\(L_i^{j*}\) denote the optimal length of the i-th PA in the m-th case (\(m \in \{1,2,3,4\}\)), with \(s^{j*}\) and \(t^{j*}\) being the corresponding optimal values of $s$ and $t$ in different cases, respectively.
			\paragraph{Case 1} The maximum values of $\Omega_{1}$ and $\Omega_{2}$ reach $1$. 
			From \eqref{varOmega1}, the maximum value $\Omega_{1} = 1$ holds if and only if $\omega_{1} \le 0$ and $\omega_{2}\ge \frac{D^{2}}{4}$, which is equivalent to the following relations: 
			\begin{align}
				%^	\Omega_{1}&=1 \Leftrightarrow \omega_{1}\le 0 , \omega_{2}\ge \frac{D^{2}}{4}\notag\\
				\begin{cases}
					AF\rho_{t}r^{1*} - d^{2} \le 0,\\
					BF\rho_{t}r^{1*} - d^{2} \ge \frac{D^{2}}{4}.
				\end{cases}
			\end{align}
			After  performing mathematical transformations on the above formulas, we obtain the range of 
			$r^{1*}$ as 
			\begin{align}
				\frac{d^{2}+ D^{2}/4}{BF\rho_{t}} \le r^{1*} \le\min\left(1,\frac{d^2}{A\rho_{t}F}\right). \label{s lower bound} 
			\end{align}
			
			From \eqref{Omega2}, $\Omega_{2}$ obtains the maximum value of $1$ if and only if $\min(\omega_{3},\omega_{4}) \ge \frac{D^2}{4}$. This is equivalent to 
			%	when $s^{1*}$ and $t^{1*}$ satisfy the following conditions:
			\begin{align}
				%	\Omega_{2}& = 1\Leftrightarrow \min(\omega_{3},\omega_{4}) \ge \frac{D^2}{4}\notag\\
				%	&=
				\min\left(\eta\left(\frac{\alpha_{1}}{\gamma_{1}}-\alpha_{2}\right),\eta\frac{\alpha_{2}}{\gamma_{2}}\right)F\rho_{t}\left(1-Fr^{1*}\right)t^{1*} - d^{2} \ge\frac{D^{2}}{4}.
			\end{align}
			We can obtain the co-relationship of $r^{1*}$ and $t^{1*}$ as
			\begin{align}
				\frac{d^{2}+ D^{2}/4}{\min(A,B)F\rho_{t}\left(1-Fr^{1*}\right)}\le t^{1*} \le1.
				\label{t_range}
			\end{align}
			Since $t^{1*} \in (0,1]$,   The upper bound of $r^{1*}$ is derived as
			\begin{align}
				(1-Fr^{1*})\ge\frac{d^{2}+D^{2}/4}{\min(A,B)F\rho_{t}}
				\Rightarrow r^{1*} \le \frac{1}{F}\left(1-\frac{d^{2}+D^{2}/4}{\min(A,B)F\rho_{t}}\right). \label{s upper bound}
			\end{align}
			Combining  \eqref{s lower bound} and \eqref{s upper bound}, the effective range of $r$ is written as
			\begin{small}
				\begin{align}
					&\frac{d^{2}+D^{2}/4}{BF\rho_{t}} \le r^{1*} \le \min\left(\frac{1}{F}\left(1-\frac{d^{2}+D^{2}/4}{\min(A,B)F\rho_{t}}\right),1,\frac{d^2}{A\rho_{t}F}    \right),\notag \\ 
				\end{align}
			\end{small}where \(r = 1\), i.e., \(L_1 = \frac{\pi}{2\kappa}\) is allowed only if \(1-Fs \neq 0\). Since \(r = \sin^2(\kappa L_1)\) is monotonically increasing for \(L_1 \in (0, \frac{\pi}{2\kappa}]\), the feasible region of \(L_1\) is obtained by inverting $r^{1*}$ via \(L_1^{1*} = \frac{1}{\kappa}\arcsin(\sqrt{r^{1*}})\).
			After fixing \(L_1^{1*}\), substituting \(L_1^{1*}\) into \eqref{t_range} and using the monotonicity of \(t = \sin^2(\kappa L_2)\), the feasible region of \(L_2^{1*}\) is derived similarly. 
			\paragraph{Case 2}Only $\Omega_{1}$ reaches its maximum of $1$. In order to obtain the minimum value of $\text{P}_{\text{sop}}$, we first analyze the monotonic relationship of \(\Omega_2\) with respect to $r^{2*}$ and $t^{2*}$ as
			\begin{align}
				\Omega_{2} \propto 
				\frac{2}{D}\sqrt{\min\left(\omega_{3},\omega_{4}\right)}  = \min\left(B,A\right) F\rho_{t}\left(1-Fr^{2*}\right)t^{2*}. \label{mon 2}
			\end{align}
			Since $\Omega_{2}$ is monotonically increasing with  $t$, we set $L_2^{2*} = \frac{\pi}{2\kappa}$. With \(L_2 = L_{2}^{2*}\), problem \(\mathcal{P}_1\) reduces to a single-variable optimization over $r^{2*}$, i.e., \(L_1^{2*}\), where the objective function is written as
			\begin{align}\label{f s}
				\min \text{P}_{\text{sop}}  \Leftrightarrow \max f(r^{2*}) = \Omega_{1}(r^{2*})\cdot \left(1-Fr^{2*}\right),
			\end{align}
			where $1-Fr$ denotes the residual power ratio in the waveguide after  PA-1 radiates power.
			When \(\Omega_1(r^{2*}) = 1\), i.e., \(r^{2*} \ge \frac{d^{2}+D^{2}/4}{BF\rho_{t}}\), \(f(r^{2*}) = 1 \cdot (1-Fr^{2*})\), which is monotonically decreasing with r. Thus, the optimal s is the minimum value that satisfies \(\Omega_1(r^{2*}) = 1\), i.e., \(r^{2*} = \frac{d^{2}+D^{2}/4}{BF\rho_{t}}\). The corresponding optimal \(L_1\) is given by
			\begin{align}
				L_{1}^{2*} = \frac{1}{\kappa}\arcsin(\sqrt{\frac{d^{2}+D^{2}/4}{BF\rho_{t}}}).
			\end{align}
			\paragraph{Case 3} Only $\Omega_{2}$ reaches its maximum of $1$.  Similarly,  the monotonicity of $\Omega_{1}$ with respect to $r^{3*}$ is analyzed as
			\begin{align}
				\Omega_{1} \propto \vert \omega_{2} - \omega_{1} \vert = \left(B-A\right)F\rho_{t}r^{3*}, \label{mon 1}
			\end{align}
			where $\omega_{1}\ge0,\omega_{2}\le\frac{D^2}{4}$, the feasible region of $r^{3*}$ is given as
			\begin{align}
				\frac{d^{2}}{AF\rho_{t}} \le r^{3*} \le \frac{d^2+D^{2}/4}{BF\rho_{t}}.
			\end{align}
			Since $\text{P}_{\text{sop}}$ is monotonicity decreasing with $r^{3*}$, the optimal value of $r^{3*}$
			is $\frac{d^2+D^{2}/4}{BF\rho_{t}}$, and the feasible region of $t$ is given as
			\begin{align}
				\frac{d^{2}+D^2/4}{\min\left(A,B\right)F\rho_{t}\left( 1-\frac{F(d^{2}+D^{2}/4)} {BF\rho_{t}}\right)}\le t^{3*}\le 1.
			\end{align}
			Through algebraic manipulations on $r^{3*}$ and $t^{3*}$, the proof of case 3 is completed.
			\paragraph{Case 4} The maximum values of $\Omega_{1}$ and $\Omega_{2}$ are less than $1$.  We consider the situation that $0<\omega_{1}<\omega_{2}<\frac{D^{2}}{4}$. Based on the monotonicity analysis of $\Omega_{1}$ and $\Omega_{2}$ presents in \eqref{mon 1} and \eqref{mon 2}, $f(r^{4*})$ is written as
			\begin{align}
				f(r^{4*})  = \frac{2}{D}\left(\sqrt{\omega_{2}} - \sqrt{\omega_{1}}\right)\cdot\Omega_{2}(r^{4*},1).
			\end{align}
			Since $B>A$, $\Omega_{2}(r,1)  = \frac{2}{D}\sqrt{\omega_{3}}$. $f(r^{4*})$ can be given by
			\begin{align}
				f(r^{4*})  =& \frac{D^{2}}{4}\left(\sqrt{BF\rho_{t}r^{4*} -d^{2}} - \sqrt{AF\rho_{t}r^{4*} - d^{2}}\right)\notag\\ &\times\sqrt{BF\rho_{t^{4*}}(1-Fr^{4*})-d^{2}}.
			\end{align}
			Maximizing \(f(r^{4*})\) is equivalent to maximizing its square, given that the range of \(f(r^{4*})\) is non-negative, i.e., \(f(r^{4*}) \geq 0\) for all s in the domain. We thus define an auxiliary function \(g(r^{4*})\) as the square of \(f(r^{4*})\), i.e., \(g(r^{4*}) = [f(r^{4*})]^2\). We
			differentiate \(g(r^{4*})\) with respect to $r^{4*}$ and set the derivative to zero as
			\begin{align}
				\frac{dg(r^{4*})}{dr^{4*}} \propto u^{'}(r^{4*})v(r^{4*}) + u(r^{4*})v^{'}(r^{4*}),
			\end{align}
			where $u(r^{4*}) =\left(\sqrt{BF\rho_{t}r^{4*} -d^{2}} - \sqrt{AF\rho_{t}r^{4*} - d^{2}}\right)^{2} $, $v(r^{4*}) = BF\rho_{t}(1-Fr^{4*})-d^{2}$, $u^{'}(r^{4*}) = \left(\sqrt{BF\rho_{t}r^{4*} -d^{2}} - \sqrt{AF\rho_{t}r^{4*} - d^{2}}\right)$
			$\times\left(\frac{BF\rho_{t}}{\sqrt{BF\rho_{t}r^{4*} -d^{2}} }-\frac{AF\rho_{t}}{\sqrt{AF\rho_{t}r^{4*} -d^{2}} }\right)$, and $v^{'}(r^{4*})   = -BF^{2}\rho_{t^{4*}}$.
			
			To obtain an approximate solution, we further assume that $\omega_{1}$ and $\omega_{2}$ are much greater than 0. Correspondingly, \(f(r^{4*})\) can be approximated as 
			\begin{align}
				f(r^{4*})\approx&\frac{4}{D^{2}}\sqrt{r^{4*}}\left(\sqrt{BF\rho_{t}} - \sqrt{AF\rho_{t}}\right)\notag\\
				&\times\sqrt{BF\rho_{t}(1-Fr^{4*})-d^{2}},
			\end{align}
			which attains its extremum when \(r^{4*} = \frac{BF\rho_{t}-d^{2}}{2BF^{2}\rho_{t}}\).  Since there is no elementary analytical solution, we first determine the feasible interval of $r^{4*}$ as $ r^{4*} \in \left( \frac{d^2}{A F \rho_t}, \min\left( \frac{d^2 + D^2/4}{B F \rho_t}, \frac{B F \rho_t - d^2}{B F \rho_t^2}, 1 \right) \right) $,  and then solve for the optimal solution using the bisection method.
			The proof is completed.
		\end{appendices}

		\bibliographystyle{IEEEtran}   %参考文献引用
		\bibliography{ref_panomasec}
		
		\vfill
	\end{document}